\documentclass[a4paper,fleqn]{cas-sc}

\usepackage{todonotes}
\usepackage[normalem]{ulem}
\usepackage{tabularx}
\usepackage{algorithm,setspace}
\usepackage{algpseudocode}
\usepackage{amsmath}
\newcounter{todocounter}

\newcommand{\brr}[1]{\stepcounter{todocounter}
  {\color{red} BRR: \thetodocounter: #1}}

\usepackage[authoryear,longnamesfirst]{natbib}
\usetikzlibrary{patterns}

\begin{document}
\let\WriteBookmarks\relax
\def\floatpagepagefraction{1}
\def\textpagefraction{.001}

\shorttitle{Transport via Backward Characteristics Tracing}    

\shortauthors{Dolence et al.}  

\title [mode = title]{A Method for Thermal Radiation Transport Using Backward Characteristic Tracing}



%

\author[1,3,4]{J. C. Dolence}[orcid=0000-0003-4353-8751]

\cormark[1]





\author[2]{H. R. Hammer}[orcid=0000-0003-0745-1771]





\author[2]{H. Park}[orcid=0000-0002-0707-9677]





\author[3]{B. Prather}[orcid=0000-0002-0393-7734]

\author[3]{B. R. Ryan}[orcid=0000-0001-8939-4461]





\author[3]{R. T. Wollaeger}[orcid=0000-0003-3265-4079]





\affiliation[1]{organization={Michigan SPARC, Los Alamos National Laboratory},
            addressline={3520 Green Ct}, 
            city={Ann Arbor},
            postcode={48105}, 
            state={MI},
            country={USA}}

\affiliation[2]{organization={T-3, Los Alamos National Laboratory},
            addressline={P.O. Box 1663}, 
            city={Los Alamos},
            postcode={87545}, 
            state={NM},
            country={USA}}

\affiliation[3]{organization={CCS-2, Los Alamos National Laboratory},
            addressline={P.O. Box 1663}, 
            city={Los Alamos},
            postcode={87545}, 
            state={NM},
            country={USA}}

\affiliation[4]{organization={Aerospace Engineering, University of Michigan},
            addressline={1320 Beal Avenue},
            city={Ann Arbor},
            postcode={48109},
            state={MI},
            country={USA}}

\cortext[1]{Corresponding author}



\begin{abstract}
Thermal radiation transport is a challenging problem in computational physics that has long been approached primarily in one of a few standard ways: approximate moment methods (for instance P$_1$ or M$_1$), implicit Monte Carlo, discrete ordinates, and long characteristics. In this work we consider the efficacy of the Method of (Long) Characteristics (MOC) applied to thermal radiation transport. Along the way we develop three major ideas: transporting MOC particles backwards in time from quadrature grids at the end of the timestep, limiting the computational cost of these backward characteristics by terminating transport once optical depths along rays become sufficiently large, and timestep-dependent closures with multigroup MOC solutions for a gray low-order system. We apply this method to a suite of standard radiation transport and radiation hydrodynamics test problems. We compare the method to several standard analytic and semi-analytic solutions, as well as implicit Monte Carlo, P$_1$, and discrete ordinates (S$_n$). We see that the method: gives excellent agreement with known results, has stability for large time steps, has the diffusion limit for large spatial cells, and achieves $\sim$20-70\% performance improvement when terminating optical depths at O(10-100) in the grey Marshak and crooked pipe problems.
However, for the Coax radiation-hydrodynamics problem, we see that MOC is approximately two to three times slower than IMC-DDMC and S$_n$ in its current implementation.
\end{abstract}




\maketitle

\section{Introduction}\label{sec:intro}

Thermal radiation transport is an important physical effect in many areas of engineering and astrophysics. 
In general, producing accurate solutions to the equations of thermal radiation transport is challenging.
This challenge is largely due to the non-linear coupling of the high-dimensional (7-D) transport equation, which evolves the radiation specific intensity $I_{\nu}\left({\bf x}, {\bf \Omega}, \nu, t\right)$ in phase space, to matter; this coupling may also be stiff due to large absorption and scattering coefficients (opacity).
This complexity often precludes analytic treatments, and numerical treatments often struggle to achieve acceptable resolutions on available hardware at practical costs.

For the better part of the last century, thermal radiation transport has usually been solved numerically in one of two ways.
For brevity, in the following discussion we exclude approximate moment methods, such as flux-limited diffusion \cite{Levermore1981}, P$_1$, or M$_1$ \cite{Levermore1984}.
The first is with the Monte Carlo method, where $I_{\nu}$ is discretized into particles with statistically sampled positions, directions, and frequencies.
These particles then evolve as if they were individual photons, transporting through space and undergoing absorption and scattering events.
This method produces a solution to the radiation transport equation that converges as $N^{-1/2}$ where $N$ is the number of particles.
While this convergence rate is advantageous for high-dimensional problems because there is no scaling with problem dimension, it is disadvantageous in that answers converge slowly as particle resolution is increased.

Monte Carlo thermal radiation transport is usually stabilized in the presence of strong energy coupling with an underlying material with the implicit Monte Carlo (IMC) method (\cite{FleckCummings1971}, and see \cite{Wollaber2016} for a review), which introduces effective scattering (absorption followed by instantaneous reemission) to mitigate over-absorption in a time step.
This method as usually implemented is actually only semi-implicit in time, although fully implicit iterative techniques exist \citep{Cleveland2018}. 
Effective scattering in IMC can be prohibitively expensive in regions of large optical depth, where many interaction events each need to be computed explicitly.
Acceleration techniques have been developed to mitigate this, primarily random walk \citep{FleckCanfield1984,Keady2017,Richers2017} and discrete diffusion Monte Carlo \citep{Gentile2001,Densmore+2007,Densmore+2012,Abdikamalov2012,Cleveland2014,smith2018,Yan2025}.
These methods are effective at accelerating diffusive calculations but retain the weak convergence rates of the Monte Carlo method and require heuristic treatments for discriminating between diffusion and transport regions of phase space~\citep{FleckCanfield1984,Densmore+2007}.

The second major numerical method for thermal radiation transport is discrete ordinates (S$_n$) or short characteristics \citep{Carlson1955,Chandrasekhar1960}.
The space of photon directions over the unit sphere is discretized with a quadrature consisting of the discrete ordinates and an associated set of weights, which take into account conservation of $4\pi$ in solid angle.
S$_n$ has ``short'' characteristics, since the intensity field is discretized in space and an auxiliary equation relating intensity on the cell interior to the cell boundary is employed.
Discontinuous Galerkin (DG) finite elements are a standard approach to preserve the asymptotic diffusion limit with S$_n$ (see, for instance, \cite{Adams1998,Adams2001,Mcclarren2008}); in this limit, the method provides a valid discretization of the equilibrium diffusion equation.
The main drawbacks of S$_n$ have historically been: ray effects \citep{Lathrop1968}, where a localized isotropic source in an optically thin medium shows spatially discrete rays; and dense memory usage, given a full intensity quadrature is stored per finite element per cell.

A transport method, Monte Carlo or deterministic, can also serve as a closure to a scheme that directly evolves moments of the photon transport equation, along with the equations for matter.
Several approaches for this exist, including high-order low-order (HOLO) (\cite{Chacon+2017,yee_stable_2017,park_asymptotic_2019}), variable Eddington tensor (VET, or variable Eddington factor, VEF; for instance \cite{Stone1992,Gnedin2001,jiang_godunov_2012,olivier_variable_2017,anistratov_dfem_2017,Anistratov2018,lou_variable_2021,Anistratov+2024}), and the second moment method \citep{Lewis1976,Olivier2025}.
In this context, the transport discretization does not necessarily need to have the spatial DG element properties outlined by \cite{Adams2001} to preserve the asymptotic diffusion limit (see, for instance \cite{Warsa2018,lou_variable_2019,Jiang2021,Anistratov+2024,Olivier2025}).

The Method of Characteristics (MOC) has recently seen renewed interest in the context of thermal radiation transport, for instance \cite{Hammer+2019}, \cite{RyanDolence2020}.
Unlike S$_n$, MOC solves for intensities along paths that traverse cells, where opacity and emissivity are discrete, but the intensity solution is exact given these approximations.
Consequently, it can leverage particle tracking infrastructure from Monte Carlo, but for intensity rays, and achieve $N^{-1}$ convergence, where $N$ is again the number of particles.
Rather than sample particle directions stochastically, here we consider MOC with discrete ordinate quadratures often used in S$_n$ methods.
MOC does not enforce energy conservation by itself (though there exist amendments to the method that do so~\cite{Hammer+2019}).
In order to enforce energy conservation, we solve low-order equations for the energy density and radiative flux with a standard second-order spatial discretization, and close these moment equations with the high-order MOC solution.
The MOC intensity particles are evolved backwards in time from spatial cell vertices, permitting evaluation of Eddington tensors and MOC flux at vertices, faces, and cell centers.
Furthermore, particles are terminated after they have tracked backwards in time beyond an optical depth threshold.

This paper is organized as follows.
In Section \ref{sec:eqns}, we present the form of the transport equation solved, along with the backward characteristics equations, the corresponding low-order system, and consistency data that couples the high and low orders.
In Section \ref{sec:num}, we discuss the discretization of the MOC and low-order solver, including a time step-dependent multigroup collapse of low-order coefficients.
In Section \ref{sec:test}, we test our reverse-time MOC implementation on a suite of standard test problems in the thermal radiation transport and radiation hydrodynamics literature.
Finally, in Section \ref{sec:conc} we outline conclusions and future work.

\section{Equations}\label{sec:eqns}

We will solve the frequency-dependent radiation transport equation with emission, absorption, isotropic elastic scattering, and material motion corrections \citep{LowrieWollaber2014}: \begin{align}
\frac{dI_{\nu}}{ds} = -\sigma_{{\rm t},\nu} I_{\nu} + \sigma_{{\rm a},\nu} {c} B_{\nu} + \frac{1}{4\pi} \sigma_{{\rm s},\nu}{c}E_{\nu} + \frac{1}{4\pi}\left( D_E + D_F\right)
\label{eqn:transport}
\end{align}
where $d/ds$ denotes the spatial derivative along the characteristic direction, $\sigma_{{\rm a},\nu}$ is the absorption opacity, $\sigma_{{\rm s},\nu}$ is the scattering opacity, $\sigma_{{\rm t},\nu} = \sigma_{{\rm a},\nu} + \sigma_{{\rm s},\nu}$ is the total opacity, $B_{\nu}(T) = \frac{2 h \nu^3}{c^3}\left(e^{h\nu/kT}-1\right)^{-1}$ is the Planck function, $T$ is material temperature, and the material motion corrections are
\begin{align}
D_E &=  -\sigma_{{\rm t},\nu} \frac{\bf v}{c} \cdot \left[{\bf F}_{\nu} - \frac{4}{3}E_{\nu} {\bf v} \right], \\
D_F &= 4 \sigma_{{\rm t},\nu} E_{\nu} {\bf \Omega} \cdot {\bf v}
\end{align}
where ${\bf v}$ is the material velocity, $E_{\nu}$ is the frequency-dependent radiation energy density, and ${\bf F}_{\nu}$ is the frequency-dependent radiation flux. 

\subsection{Backwards-in-time transport}

It is straightforward to integrate Equation \ref{eqn:transport} {\it backwards} in time from a final time $t_{\rm f}$ while still recovering the final intensity $I_{\nu}\left( t_{\rm f}\right)$ \citep{DolenceThesis}. Begin from the formal solution of the transfer equation,
\begin{align}
I(\tau) = I_0 e^{-(\tau - \tau_0)} + \int_{\tau_0}^{\tau} d\tau' S(\tau') e^{-(\tau' - \tau_0)}
\label{eqn:formalsoln}
\end{align} 
where
\begin{align}
\tau = \int_0^s \sigma(s') ds'.
\end{align}
The boundary condition $I_0$ is already related to the final state $I(\tau)$ by the total optical depth $\tau$. There is also no directionality to the integrals over the opacity $\sigma$ and the source term $S$; we can equivalently solve over a backward-in-time characteristic $\eta' = s-s'$ with $s_0 > s$,
\begin{align}
\tau &= \int_0^s \sigma(s') ds' = \int_0^s\sigma(\eta')d\eta' \;\;,\\
I(\tau_0) &= I(\tau) e^{-(\tau_0 - \tau)} + \int_{\tau}^{\tau_0} d\tau' S(\tau') e^{-(\tau_0 - \tau')} \;\;.
\label{eqn:backwardsformalsoln}
\end{align}
Hence, the procedure for integrating optical depth along characteristics should be unchanged, but here $\tau_0$ is now optical depth from a further point in the future (a minimal point in $\eta'$).

The motivation for solving the intensity backwards in time is closely related to the motivation for adjoint transport methods, where the solution corresponds to a level of importance of a spatial point to the response at a different detector point \citep{Bell1970}.
Here, we seek an intensity solution at a particular (response) point and time, and forcing characteristics to end at exactly at the point and time is accomplished by evolving backwards in time from this space-time point, as in time-dependent adjoint transport.
This also establishes a numerical discretization with appealing properties at the end of each timestep, for instance a complete quadrature at particular points when performing integrals over momentum space.
Such a backwards-in-time approach to transport also allows us to halt work early beyond an optical depth that will mask all sources further back in time. Note that $\sigma$ and $\tau$ are assumed to be time-independent here; this would likely first become an issue if one were to include the full scattering kernel in $S$, but as seen we have already approximated the scattering source in terms of the radiation energy density $E_{\nu}$. We discuss our numerical discretization of Equation \ref{eqn:backwardsformalsoln} in Section \ref{sec:moc_disc}.

\subsection{Moment equations}

We will also solve the gray moment equations consistent with Equation \ref{eqn:transport}:
\begin{align}
\frac{\partial E}{\partial t} + \nabla \cdot {\bf F} &= \sigma_{\rm P} {c} B -\sigma_{\rm a} c E - \sigma_{\rm t} \frac{\bf v}{c} \cdot {\bf F} + \frac{4}{3}\sigma_{\rm t}\frac{v^2}{c} E\label{eqn:mom0}\\
{\frac{1}{c}}\frac{\partial {\bf F}}{\partial t} + \nabla \cdot {c}\mathbb{P} &= -\boldsymbol{\sigma}_{\rm t}:{\bf F}  + \frac{4}{3} \sigma_{\rm t} {\bf v} E \label{eqn:mom1}
\end{align}
where $B=aT^4$ is the integral of $B_{\nu}$ over solid angle and frequency, $\boldsymbol{\sigma}_{\rm t}:{\bf F}$ is an element-wise product of vectors $\boldsymbol{\sigma}_{\rm t}$ and ${\bf F}$, and the opacities without $\nu$ subscripts are appropriately frequency-averaged:
\begin{align}
    \sigma_P &= \frac{4\pi}{cB}\int d\nu B_{\nu}\sigma_{a,\nu} \;\;, \\
    \sigma_a &= \frac{1}{cE}\int d\Omega d\nu I_{\nu}\sigma_{a,\nu} \;\;,\\
    \sigma_{t,j} &= \frac{1}{F_j}\int d\Omega d\nu \Omega_jI_{\nu}\sigma_{t,\nu} \;\;,
\end{align}
where subscript $j$ indicates a component of the vector.
The moments of the radiation intensity
\begin{align}
E &= \frac{1}{c}\int d\Omega d\nu I_{\nu} \;\;, \\
\mathbf{F} &= \int d\Omega d\nu {\bf \Omega} I_{\nu} \;\;, \\
\mathbb{P} &= \frac{1}{c}\int d\Omega d\nu {\bf \Omega} \otimes {\bf \Omega} I_{\nu} \;\;,
\end{align}
are the radiation energy density, flux, and pressure tensor, respectively.

Note that Equations \ref{eqn:mom0} and \ref{eqn:mom1}, which provide a natural avenue to coupling with a hydrodynamical material representation, due to representing the same energy and momentum quantities, require both frequency-averaged opacities and pressure tensors. The central design of our method is to use a solution of Equation \ref{eqn:transport} {\it only} to provide this closure; the moment representation will be the true representation of the radiation energy and momentum. This approach provides us considerable freedom in how we solve Equation \ref{eqn:transport}; for example, we can without difficulty adopt non-conservative methods, such as MOC.

\subsection{Consistency}

We evolve both the high-order $I_{\nu}$ and the low-order $E$ and ${\bf F}$ as dynamical quantities. However, integrals over $I_{\nu}$ will produce competing representations of $E$ and ${\bf F}$ that will in general disagree with the low-order system. To distinguish these quantities, we indicate the former as $E_{\rm HO}$, ${\bf F}_{\rm HO}$ and the latter as $E_{\rm LO}$, ${\bf F}_{\rm LO}$.

Our overall design is to treat the low-order sector as the true representation of the energy and momentum of the radiation, and have the high-order sector provide guidance to the low-order sector during evolution through normalized pressure tensors
\begin{align}
\mathbb{P} = E_{\rm LO} \frac{\int d\Omega d\nu {\bf \Omega} \otimes {\bf \Omega} I_{\nu}}{\int d\Omega d\nu  I_{\nu}}
\end{align}
and averaged opacities
\begin{align}
    \sigma_{\rm a} &= \frac{\int d\Omega d\nu \sigma_{{\rm a},\nu} I_{\nu}}{\int d\Omega d\nu  I_{\nu}} \;\;, \\
    \sigma_{\rm s} &= \frac{\int d\Omega d\nu \sigma_{{\rm s},\nu} I_{\nu}}{\int d\Omega d\nu  I_{\nu}} \;\;.
\end{align}
The gray average opacities are used in the radiation energy density, material coupling, and low-order material motion corrections.
In multigroup problems, for the flux equation spatial stencil, we use two additional averages of HO data, representing flux- and pressure tensor-weighted harmonic averages of the number of mean-free times.
These additional averages are discussed in Section \ref{sec:moments}.

In turn, frequency dependent low-order variables that appear in scattering and moving material correction sources in Equation \ref{eqn:transport} are constructed by combining the spectral shape of the high-order system and the normalization of the low-order system:
\begin{align}
    E_{\nu} &= \frac{E_{{\rm HO},\nu}}{E_{\rm HO}}E_{\rm LO} \;\;, \\
    {\bf F}_{\nu} &= \frac{{\bf F}_{{\rm HO},\nu}}{{\bf F}_{\rm HO}}{\bf F}_{\rm LO} \;\;.
\end{align}

We do not have artificial consistency terms designed to drive the high-order solution towards the low-order solution as in HOLO methods \citep{Chacon+2017}. Physical source terms will drive both high-order and low-order sectors to a consistent equilibrium, but remaining inconsistencies are simply a manifestation of truncation error (in both spatial and temporal discretizations) and do not themselves drive instabilities in the method, as the high order sector provides only normalized quantities (averaged opacities, normalized tensors) to the low-order sector during evolution.

\subsection{Boundary conditions}

For the numerical tests, we have implemented vacuum, reflective, and isotropic surface source boundary conditions.
Apart from the effect of the boundary condition on Eddington tensors, vacuum and reflective boundary conditions are treated separately in the HO and LO systems.
For the surface source, we use the Miften-Larson boundary condition \citep{Miften1993} when $E_{HO}>0$ and fall back to a fully HO boundary condition if $E_{HO}=0$:
\begin{align}
    {\bf F}_{LO}\cdot {\bf n} = \begin{cases}
        \left(F_{HO}^+ + F_{HO}^-\right)\frac{E_{LO}}{E_{HO}} - 2F_{HO}^- \;\;,\;\; E_{HO} > 0 \;\;, \\
        F_{HO}^+ - F_{HO}^- \;\;,\;\; E_{HO} = 0 \;\;,
    \end{cases}
\end{align}
where ${\bf n}$ is a normal vector pointing outward from the spatial domain and $F_{HO}^+$ ($F_{HO}^-)$ is the partial flux out of (into) the domain along ${\bf n}$.
If $E_{LO}=E_{HO}$, it is apparent that the Miften-Larsen flux reduces to the fully HO flux.
If the HO intensity is linearly anisotropic with respect to the component of direction along ${\bf n}$, then the Miften-Larsen boundary condition reduces to the standard Marshak boundary condition \citep{Miften1993}.

\section{Numerical discretization}\label{sec:num}

We now describe our numerical method for solving Equations \ref{eqn:transport}, \ref{eqn:mom0}, and \ref{eqn:mom1} simultaneously in high-order and low-order sectors. The high-order radiation transport equation is solved with a MOC approach in which $I_{\nu}$ is evolved deterministically along characteristics, and the low-order moment equations are evolved with a 2nd order finite volume and finite difference method. The two sectors are linked via flux and pressure tensor coefficients, and emission, absorption, scattering, and moving material correction source terms. 

The MOC transport updates intensity along rays assuming piecewise-constant emission sources in each spatial cell, discussed in detail in Section~\ref{sec:moc_disc}.
We expect the LO solver to enforce the diffusion limit, given the MOC-LO coupling scheme we implement.
However, linear finite elements should still lower spatial truncation error.  
The application of higher order spatial elements is left as future work.

We have developed our implementation inside the Jayenne transport library \citep{Thompson+2021}, which provides particle and mesh support for IMC that can be leveraged for MOC.
Jayenne includes infrastructure for cell-based adaptive mesh refinement (AMR) and MPI-based domain decomposition.
Details for AMR and domain-decomposition associated specifically with MOC are discussed in Section~\ref{sec:dlbamr}.

\subsection{Method of Characteristics}
\label{sec:moc_disc}

Here, we elaborate the algorithm for integrating optical depth and intensity along particle objects that are tracked backward in time; this is a main feature of this work, as it permits optical depth-based optimization.
Given that intensity exponentially attenuates along a ray when moving forward in time, one might construct a scheme where absorption causes intensity to exponentially grow backward in time.
However, positive exponents may cause floating point overflow and are in fact unnecessary, as will be shown here.

We lay down samples on a quadrature grid at each node at the end of each timestep. We position particles on nodes rather than cell centers for two reasons. First, interpolation between support points at the beginning of each timestep from the previous timestep's updated samples is simplified, especially on arbitrarily refined meshes. Second, interpolation is always internal to a cell which prevents interpolation error from dominating weak fluxes in the diffusive regime.
Each particle stores a full spectrum of multigroup intensity and optical depths (see \cite{Morgan+2021} for an example with Monte Carlo).

\begin{figure}[h!]
    \centering
    \begin{minipage}{0.45\textwidth}
        \centering
        \resizebox{.6\textwidth}{!}{
\begin{tikzpicture}

    \node[text=black] at (1.5,3.5) {$t = t_{\rm f}$};
    \foreach \x in {0,1,2,3} {
        \foreach \y in {0,1,2,3} {
            \filldraw[black] (\x,\y) circle (1.5pt); 
            \ifnum\x<3
                \draw[black] (\x,\y) -- (\x+1,\y); 
            \fi
            \ifnum\y<3
                \draw[black] (\x,\y) -- (\x,\y+1); 
            \fi
        }
    }


    \draw[->, red, thick] (1,1) -- (1.25,1.25);
    \draw[->, cyan, thick] (1,1) -- (1.25,0.75);
    \draw[->, orange, thick] (1,1) -- (0.75,1.25);
    \draw[->, violet, thick] (1,1) -- (0.75,0.75);

\end{tikzpicture}
}
\end{minipage}
\begin{minipage}{0.45\textwidth}
        \centering

        \resizebox{.6\textwidth}{!}{
\begin{tikzpicture}
    \node[text=black] at (1.5,3.5) {$t = t_{\rm 0}$};
        \foreach \x in {0,1,2,3} {
            \foreach \y in {0,1,2,3} {
                \filldraw[black] (\x,\y) circle (1.5pt); 
                \ifnum\x<3
                    \draw[black] (\x,\y) -- (\x+1,\y); 
                \fi
                \ifnum\y<3
                    \draw[black] (\x,\y) -- (\x,\y+1); 
                \fi
            }
        }


        \draw[dashed, red, thick] (.5,.5) -- (1.,1.);
        \draw[dashed, violet, thick] (1.5,1.5) -- (1,1);
        \draw[dashed, cyan, thick] (.5,1.5) -- (1,1);
        \draw[dashed, orange, thick] (1.5,.5) -- (1,1);

        \draw[->, red, thick] (.5,.5) -- (.75,.75);
        \draw[->, violet, thick] (1.5,1.5) -- (1.25,1.25);
        \draw[->, cyan, thick] (.5,1.5) -- (.75,1.25);
        \draw[->, orange, thick] (1.5,.5) -- (1.25,.75);

    \end{tikzpicture}
    }
    \end{minipage}

    \caption{A 2D mesh with MOC samples shown on their quadrature grid at the end of the cycle ($t_{\rm f}$; left) and beginning of cycle ($t_0$; right)  after transport backwards in time. Directions of arrows indicate the movement of the particle going {\it forwards} in time, the opposite direction to which particles are actually moved during numerical updates. At the end of the cycle, the samples are colocated at a vertex and lie on a prescribed quadrature grid, allowing for fast and accurate integrations over solid angle.}
    \label{fig:moc_transport}
\end{figure}
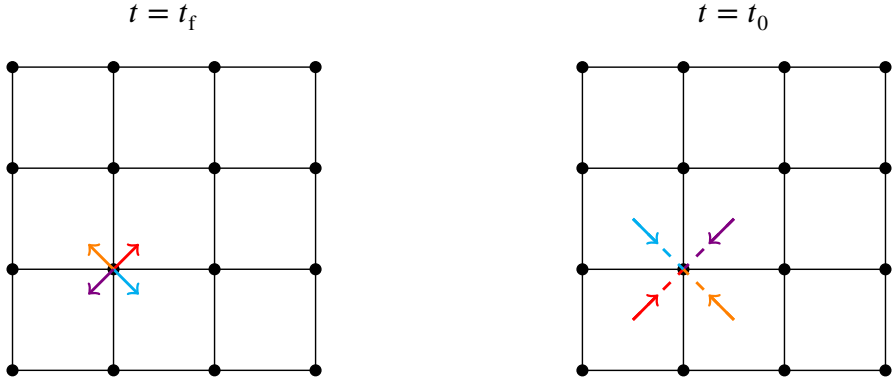

Figure \ref{fig:moc_transport} illustrates the initial distribution of samples and the backwards-in-time transport to their positions at the beginning of the timestep. In general, this set of samples will not form quadrature grids of colocated particles at the beginning of the timestep due to mesh refinement, non-Cartesian coordinate systems, arbitrary simulation timesteps, and differing angles of propagation of the samples. 

Equation \ref{eqn:backwardsformalsoln} contains two terms that contribute to the final time intensity $I_{\nu}\left( t_{\rm f} \right)$: a boundary term due to the initial intensity attenuated by the total optical depth, and the integral of the source function along the pathlength, continuously attenuated by the pathlength-dependent intervening optical depth. We can evaluate both these terms in one traversal along each trajectory from $t_{\rm f}$ to $t_0$ by splitting the source term integral up into discrete steps while also accumulating the total optical depth $\tau$. This can be done to arbitrary spatial order. Consider a piecewise constant discretization of $S$. For a step from point $i$ to $i+1$:
\begin{align}
\int_{\tau_i}^{\tau_{i+1}} S_i e^{-\left( \tau - \tau' \right)} d\tau' &= S_i e^{-\tau} \left( e^{\tau_{i+1}} - e^{\tau_i} \right) \\
&= S_i e^{-\left(\tau - \tau_{i+1}\right)} \left( 1 - e^{-\left(\tau_{i+1} - \tau_i\right)} \right) \;\;.
\end{align}
Defining $\Delta\tau_i = \tau_{i+1} - \tau_i$ and $\bar{\tau}_i = \tau - \tau_{i+1}$:
\begin{align}
\int_{\tau_i}^{\tau_{i+1}} S_i e^{-\left( \tau - \tau' \right)} d\tau' = S_i e^{-\bar{\tau}_i} \left( 1 - e^{-\Delta \tau_i} \right) \;\;.
\end{align}

Given $N+1$ spatial points along a characteristic with piecewise constant sources
and opacities along the segments between the points, the source integral and optical depth integrals can be partitioned into $N$ segments.
Then $\tau = \tau_{N+1} = \sum_{i=N}^1\Delta\tau_i$, $\tau_1=0$, and the intensity solution along a characteristic is
\begin{align}
\bar{\tau}_i &= \begin{cases}
\displaystyle\sum_{j=N}^i \Delta \tau_j \;\;&,\;\; i\leq N \\ 
\displaystyle \;\;\; 0 \;\;&,\;\;i=N+1\end{cases} \\
I_{N+1} &= I_0 e^{-\tau} + \sum_{i=N}^1 e^{-\bar{\tau}_{i+1}} \left( 1 - e^{-\Delta \tau_i} \right) S_{i} \label{eq:moc_soln}\;\;.
\end{align}

The $N+1$ spatial points represent cell edge intersections with the particle characteristic starting at a cell vertex given by the point $N+1$ and traveling backwards in time to point 1 (which may be on the domain boundary or interior of a spatial cell).
Hence, $S_i$ and $\Delta\tau_i$ are the value of the source and the amount of optical depth traversed over the spatial cell corresponding to step $i$ (between points $i+1$ and $i$).
We evaluate the optical depth $\bar{\tau}_i$ along a characteristic simply by summing $\Delta\tau_i$ contribution from each cell the particle traverses, which is the same approach as tracing the characteristic forward in time.
For each particle, we evaluate an auxiliary intensity attached to the particle as it traverses step $j$ (going from $N$ down to 1):
\begin{align}
\tilde{I}_{N+1}^{(j)} = \sum_{i=N}^j e^{-\bar{\tau}_{i+1}} \left( 1 - e^{-\Delta \tau_i} \right) S_{i} \;\;.
\end{align}
Thus a recursion gives the update to the particle $\tilde{I}$ value after step $j-1$:
\begin{align}
\tilde{I}_{N+1}^{(j-1)} = \tilde{I}_{N+1}^{(j)} + e^{-\bar{\tau}_j}(1-e^{\Delta\tau_{j-1}})S_{j-1} \;\;, \label{eq:moc_recurse}
\end{align}
where the added term is simply the intensity from the source in step $j-1$ attenuated by the optical depth of the path from point $j$ back to point $N+1$.
The particle terminates at $j=1$, where $I_0$ from the initial or boundary condition is added to the final auxiliary intensity to get the true intensity of the particle at $N+1$: 
\begin{align}
I_{N+1} = I_0e^{-\tau_{N+1}} + \tilde{I}_{N+1}^{(1)} \;\;. \label{eq:moc_apply_ic}
\end{align}




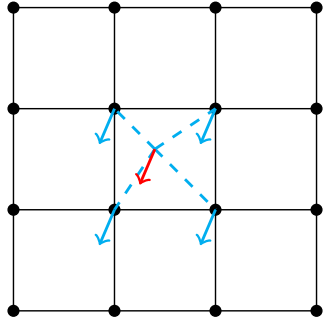
\begin{figure}[h!]
    \centering
    \begin{minipage}{0.45\textwidth}
        \centering
        \resizebox{.6\textwidth}{!}{
\begin{tikzpicture}
        \foreach \x in {0,1,2,3} {
            \foreach \y in {0,1,2,3} {
                \filldraw[black] (\x,\y) circle (1.5pt); 
                \ifnum\x<3
                    \draw[black] (\x,\y) -- (\x+1,\y); 
                \fi
                \ifnum\y<3
                    \draw[black] (\x,\y) -- (\x,\y+1); 
                \fi
            }
        }


        \draw[dashed, cyan, thick] (1.4,1.6) -- (1,1);
        \draw[dashed, cyan, thick] (1.4,1.6) -- (2,1);
        \draw[dashed, cyan, thick] (1.4,1.6) -- (2,2);
        \draw[dashed, cyan, thick] (1.4,1.6) -- (1,2);


        \draw[->, red, thick] (1.4,1.6) -- (1.25,1.25);

        \draw[->, cyan, thick] (1,1) -- (1.35-.5,1.15-.5);
        \draw[->, cyan, thick] (1,2) -- (1.35-.5,1.15+.5);
        \draw[->, cyan, thick] (2,2) -- (1.35+.5,1.15+.5);
        \draw[->, cyan, thick] (2,1) -- (1.35+.5,1.15-.5);

    \end{tikzpicture}
    }
    \end{minipage}

    \caption{Illustration of MOC particle intensities being interpolated from previous end-of-timestep support point particles (cyan) with the same directions at vertices to current beginning-of-timestep position to which current MOC particle (red) was transported backwards in time to. Note that this algorithm generalizes to adaptive mesh refinement.}
    \label{fig:moc_interp}
\end{figure}

We note that the particle location after tracing the characteristic is at point 1, but it has an intensity value for the end point $N+1$, taken to be the end of the time step ($t_f$).
If the particle is stopped at the beginning of the time step on the interior of the spatial domain, we interpolate the previous intensity in the direction of the particle to the particle (point 1) location, as depicted in Figure \ref{fig:moc_interp}, giving the $I_0$ value (we employ a bilinear interpolation in 2D).
In curvilinear coordinates, the direction components change along a straight path, so the particle direction at point 1 is no longer in the ordinate set of the quadrature, if the quadrature is defined in the curvilinear coordinate system.
Consequently, for our implementation in cylindrical geometry we interpolate in both space and in radial direction component (the direction component along the symmetry axis is preserved along the characteristic, so no interpolation is required).
After the evaluation of $I_0$, the particles are returned to their original future points (for use as $I_0$ in the next time step).
This return is trivial: each particle stores the original (future) vertex index and the originating MPI rank of the original vertex.

For our optical depth-based termination of the particle transport along the characteristic, if the threshold $\tau_{\rm th}$ is reached prior to the initial or boundary condition, we no longer add the contribution of $I_0e^{-\tau}$ under the assumption that it is a negligibly small compared to $\tilde{I}_{N+1}^{(j)}$, where $j$ is the step $\bar{\tau}_j > \tau_{\rm th}$.
Consequently, the particle is communicated back to point $N+1$ with a final intensity $I_{N+1} = \tilde{I}_{N+1}^{(j)}$.
For multigroup particle intensities, the particle is only terminated after the optical depths in all groups have surpassed $\tau_{\rm th}$.

\subsection{Moments}
\label{sec:moments}

We now examine the discretization of the LO system of equations, which is the next main development for our MOC-LO formulation.
One high-level goal is to form a MOC-LO coupling, or MOC closure for the LO system, that is as simple or nearly as simple to implement as in a fully gray VET formulation.
In particular, we intend to avoid introducing additional equations outside of closure terms found in a typical fully gray VET method.
Another high-level goal is to obtain a closure that furnishes accurate spatial dependence of intensity in material with opacities that have large variability across frequency, which is non-trivial.

The moment equations \ref{eqn:mom0} and \ref{eqn:mom1} are discretized onto the mesh with $E$ at cell-centers and fluxes dotted into face normals $F \cdot n$ located at cell faces.
This discretization naturally recovers Fick's law-like asymptotic behavior when optical depths become large.
The multifrequency formulation we present here recovers a standard gray VET discretization for one frequency group, thus we do not need to consider a separate gray formulation.
It is well known that care must be taken when integrating over frequency to obtain the LO system of equations; depending on the integration the system can be numerically unstable (see, for instance, \cite{yee_stable_2017}).

Our multifrequency treatment is multigroup in the MOC transport, where opacity is assumed to be integrated over group intervals.
For opacity that is highly variable in frequency, we find that integrating opacity over groups then discretizing in time may cause the LO solver flux stencil to fail to match the HO flux.
Integrating the multigroup form of the LO system of equations can result in a flux-weighted arithmetically averaged $\sigma_t$, whereas a harmonic weighting may be more accurate (for instance, the standard Rosseland average).
However, an arithmetic average of $\sigma_t$ may be more appropriate for the material motion correction terms.

Error associated with collapsing multigroup to gray may be remedied either by adding a consistency term \citep{Wollaber+2017} or a middle-order multigroup angularly integrated system (see, for instance,~\cite{Anistratov+2024} and references therein).
Stability can be obtained using Rosseland averaging or flux-weighted arithmetic averaging \citep{yee_stable_2017}.
We adopt a distinct approach, where we derive LO coefficients from the HO data by consistently time-discretizing the multigroup moment equations formed from the HO data, and summing the result over groups.
We may focus on the 1st moment equation in particular, and can neglect material motion correction terms which will be operator-split (Section \ref{sec:mmc}).
Performing a time-implicit finite difference along one dimension,
\begin{equation}
    F_{g,i+1/2}^{(n+1)} = \frac{1}{1 + c\Delta t_n\sigma_{t,g,i+1/2}}F_{g,i+1/2}^{(n)}
    - \frac{1}{1 + c\Delta t_n\sigma_{t,g,i+1/2}}\frac{c^2\Delta t_n}{\Delta x_{i+1/2}}
    \left(\mathbb{f}_{xx,g,i+1} E_{g,i+1}^{(n+1)} - \mathbb{f}_{xx,g,i} E_{g,i}^{(n+1)}\right) \label{eq:radF_LO}
    \;\;,
\end{equation}
where superscript $n$ is time step, subscripts $i$ and $g$ are spatial cell and group index, respectively, $\Delta t_n$ is the time step duration, $\sigma_{t,g,i+1/2}$ is the total opacity per group at face $i+1/2$, $\Delta x_{i+1/2}$ is the distance between cell centers, and $\mathbb{f}_{g,i}=\mathbb{P}_{g,i}/E_{g,i}$ is the Eddington tensor per group per cell.
For any face $j$, we calculate $\sigma_{t,g,j}$ as the harmonic average of the values immediately below (-) and above (+) the face,
\begin{equation}
    \sigma_{t,g,j} = \frac{2}{\displaystyle\frac{1}{\sigma_{t,g,j}^-} + \frac{1}{\sigma_{t,g,j}^+}} \;\;,
\end{equation}
which favors smaller opacity at faces with sharp changes in opacity.
Use of a harmonic average is a choice we have made that gives the best results for our suite of numerical tests.
At an interface between a low- and high-opacity region, the average will more closely match the low opacity, so flux from the low-opacity region at the interface will use low opacity in the stencil; this furnishes an effect similar to upwinding the opacity.
One may also note that we do not use $\Delta x_i$ as weights in the average, which is another choice: weighting the reciprocal opacities by $1/\Delta x_i$ gives an optical thickness-based average, favoring the cell with both small width and opacity, whereas weighting the reciprocal opacities directly by $\Delta x_i$ favors the cell with large width and small opacity.
We do not have an argument against either weighting, so have opted for neither (or both, since $1 = \Delta x_i / \Delta x_i$).

In Eq.~\eqref{eq:radF_LO}, we have neglected off-diagonal terms proportional to $\mathbb{f}_{xy}$ as they are discussed later and do not complicate the diagonal derivation here.
A gray radiative flux equation can be constructed simply by summing Eq.\ \ref{eq:radF_LO} over groups,
\begin{equation}
    F_{i+1/2}^{(n+1)} = \sum_g\frac{F_{g,i+1/2}^{(n)}}{1 + c\Delta t_n\sigma_{t,g,i+1/2}}
    - \frac{c^2\Delta t_n}{\Delta x_{i+1/2}}
    \left(\sum_g\frac{\mathbb{f}_{xx,g,i+1} E_{g,i+1}^{(n+1)}}{1 + c\Delta t_n\sigma_{t,g,i+1/2}}
    - \sum_g\frac{\mathbb{f}_{xx,g,i} E_{g,i}^{(n+1)}}{1 + c\Delta t_n\sigma_{t,g,i+1/2}}\right)
    \;\;.
\end{equation}
Setting
\begin{subequations}
    \label{eq:gcfdt_diag}
    \begin{gather}
        A_{i+1/2} = \frac{1}{F_{HO,i+1/2,g}^{(n)}}\sum_g\frac{F_{HO,g,i+1/2}^{(n)}}{1 + c\Delta t_n\sigma_{t,g,i+1/2}}
        \;\;, \\
        \mathbb{B}_{xx,i}^{\pm} = \frac{1}{\mathbb{f}_{xx,i}E_{HO,i}^{(n+1)}}
        \sum_g\frac{\mathbb{f}_{xx,g,i} E_{HO,g,i}^{(n+1)}}{1 + c\Delta t_n\sigma_{t,g,i\pm1/2}}
        \;\;,\\
        \mathbb{f}_{xx,i} = \sum_g \mathbb{f}_{xx,g,i}\;\;,
    \end{gather}
\end{subequations}
we may rewrite the gray flux equation as
\begin{equation}
    F_{i+1/2}^{(n+1)} = A_{i+1/2}F_{i+1/2}^{(n)}
    - \frac{c^2\Delta t_n}{\Delta x_{i+1/2}}
    \left(\mathbb{B}_{xx,i+1}^{-}\mathbb{f}_{xx,i}E_{i+1}^{(n+1)} 
    - \mathbb{B}_{xx,i+1}^{+}\mathbb{f}_{xx,i}E_{i}^{(n+1)}\right)
    \;\;.
\end{equation}
In order to formulate the closure for the LO sector, we evaluate $A_{i+1/2}$ and $\mathbb{B}_i$
with the HO (MOC) data; subscripting the terms in the equation,
\begin{equation}
    F_{LO,i+1/2}^{(n+1)} = A_{i+1/2}F_{LO,i+1/2}^{(n)}
    - \frac{c^2\Delta t_n}{\Delta x_{i+1/2}}
    \left(\mathbb{B}_{xx,i+1}^{-}\mathbb{f}_{xx,i+1}E_{LO,i+1}^{(n+1)} 
    - \mathbb{B}_{xx,i+1}^{+}\mathbb{f}_{xx,i}E_{LO,i}^{(n+1)}\right)
    \;\;.
\end{equation}
We note that these $A$- and $\mathbb{B}$-terms are merely two types of weighted averages of
the coefficient resulting from time discretization, $1/(1+c\Delta t_n\sigma_t)$, hence are
also merely types of time step-dependent harmonic averages of $\sigma_t$.

For a single group, this formulation of the flux equation reduces to a typical variable Eddington
tensor approach to treating the LO.
For multiple groups, assuming $c\Delta t_n\sigma_{t,g,i+1/2}$ is many mean-free times and further
assuming $\mathbb{f}_{xx,g,i} \approx \mathbb{f}_{xx,g,i+1} \approx 1/3$, the $\mathbb{B}$-terms
reduce to
\begin{equation}
    \mathbb{B}_{xx,i}^{\pm} = \frac{1}{c\Delta t_n E_{HO,i}}\sum_g\frac{E_{HO,g,i}}{\sigma_{t,g,i\pm1/2}} 
    = \frac{1}{c\Delta t_n\sigma_{R,i}^{\pm}}\;\;.
\end{equation}
where we have introduced $\sigma_{R,i}^{\pm}$, the radiation energy density-weighted harmonic 
group average of the total face opacity, at the lower (-) or higher (+) face along the dimension.
The flux equation is then
\begin{equation}
    F_{LO,i+1/2}^{(n+1)} \approx -\frac{c}{3\Delta x_{i+1/2}}
    \left(\frac{1}{\sigma_{R,HO,i+1}^{-}}E_{LO,i+1}^{(n+1)} 
    - \frac{1}{\sigma_{R,HO,i}^{+}}E_{LO,i}^{(n+1)}\right)
    \;\;,
\end{equation}
which is a valid discretization of Fick's Law.
Consequently, we expect this approach to obtain the diffusion limit even when the HO equation itself
does not, so long as the above assumptions that furnish Fick's Law are satisfied.
We note that this difference between HO and LO behavior may incur inconsistency between HO and LO
radiation energy density solutions, but for optically thick media with low absorption-emission time
scales the discrepancy should be regulated by the use of the LO temperature to evaluate the HO emissivity.

We treat the off-diagonal Eddington tensor (``cross'') term contributions to the 1st moment equation as purely HO, which implies these terms are sources in the LO system of equations.
These cross terms then vanish in the Jacobian matrix formed to solve the equations.
Following a similar procedure as above, isolating the cross term with the time derivative,
\begin{equation}
    F_{LO,j}^{(n+1)} = -\frac{c^2\Delta t_n}{\Delta y_j}
    \left(\mathbb{X}_{xy,v_{j,2}} - \mathbb{X}_{xy,v_{j,1}}\right) \;\;,
\end{equation}
where $j$ is now the index of the cell face with normal in the $x$-direction (replacing $i+1/2$ from above), $v_{j,1}$ and $v_{j,2}$ are the indices of the lower and upper nodes bounding face $j$, and
\begin{equation}
    \label{eq:gcfdt_cros}
    \mathbb{X}_{xy,v_{j,k}} = \sum_g\frac{\mathbb{f}_{xy,g,v_{j,k}} E_{HO,g,v_{j,k}}^{(n+1)}}{1 + c\Delta t_n\sigma_{t,g,j}} \;\;,\;\; k\in\{1,2\} \;\;.
\end{equation}
These cross term equations leverage the MOC solution of intensity at vertices (hence leverage the inverse-time transport), so do not require reconstruction of HO data from cells or cell corners to vertices.
However the $\mathbb{X}$-terms are topologically distinct from the $\mathbb{B}$-terms: for 2D rectangular cells there are 4 $\mathbb{B}$-term scalars per cell and 8 $\mathbb{X}$-term scalars per cell.

In cylindrical coordinates the radiative flux equations acquire two extra "geometric" source terms. In particular, $r-$ and $z-$flux equation becomes: 
\begin{align}
    \frac{1}{c}\frac{\partial F_r}{\partial t} +  \frac{\partial \left(\mathbb{f}_{rr} cE\right)}{\partial r} + \frac{\partial \left(\mathbb{f}_{rz}cE\right)}{\partial z} + \frac{\mathbb{f}_{rr}+\mathbb{f}_{\phi\phi}}{r} cE + \sigma_t F_r = 0,\\
    \frac{1}{c} \frac{\partial F_{z}}{\partial t}+  \frac{\partial \left(\mathbb{f}_{zr} cE\right)}{\partial r} + \frac{\partial \left(\mathbb{f}_{zz} cE\right)}{\partial z} + \frac{\mathbb{f}_{zr}}{r}cE + \sigma_t F_z = 0
\end{align}
 Here, the terms, $\frac{\mathbb{f}_{rr}+\mathbb{f}_{\phi\phi}}{r} cE$ and $ \frac{\mathbb{f}_{zr}}{r}cE $ arise purely from the divergence operator in cylindrical geometry. In our discrete algorithm, we approximate these geometric terms using the gray Eddington tensor evaluated on each face and the corresponding face-averaged low-order radiation energy density, $E_{LO,ij}$. That is a face $ij$ located at radius $r_{ij}$ we write
\begin{align}
\frac{\mathbb{f}_{rr} + \mathbb{f}_{\phi\phi}}{r}  cE&\approx \frac{\mathbb{f}_{rr,ij}+\mathbb{f}_{\phi\phi,ij}}{r_{ij}} cE_{LO,ij},\\
\frac{\mathbb{f}_{rz}}{r} cE &\approx   \frac{\mathbb{f}_{rz,ij}}{r_{ij}} cE_{LO,ij}
\end{align}

We emphasize our main motivation for the $A$-, $\mathbb{B}$- and $\mathbb{X}$-terms is their effect as harmonic averages over group of the total face opacity, $\sigma_g$.
Numerically demonstrated in Section~\ref{sec:test}, we see that combination of these terms and a simple face-centered geometric terms permit the LO gray flux to emulate the HO grey flux when the HO flux has significant contributions from optically thin groups.

\subsection{Material motion corrections}
\label{sec:mmc}
Material motion correction (MMC) terms are treated slightly differently in the low-order (LO) and high-order (HO) sectors of the algorithm. In the LO system, MMC terms are evaluated explicitly, while in the HO system they are handled implicitly. Both LO and HO systems utilize the LO radiation energy density $E_{LO}$ and flux $F_{LO}$.

In the LO sector, we employ an operator-splitting approach. First, the explicit MMC terms are applied to obtain intermediate quantities:
\begin{align}
E^*_{LO,i}  &= E^{(n)}_{LO,i} - \sigma_{\rm t} \Delta t\frac{\bf v}{c} \cdot {\bf F}^{(n)}_{LO,i} + \frac{4}{3}\sigma_{\rm t}\Delta t\frac{v^2}{c} E^{(n)}_{LO,i}\label{eqn:lommc0}\\
{\bf F}^{*}_{LO,ij} &= {\bf F}^{(n)}_{LO,ij} + \frac{4}{3} \sigma_{\rm t} c\Delta t{\bf v} E^{(n)}_{LO,ij} \label{eqn:lommc1}
\end{align}
The remaining LO system is then advanced implicitly using:
\begin{align}
\frac{E^{(n+1)}_{LO}-E^{*}_{LO}}{\Delta t} + \nabla \cdot {\bf F}^{(n+1)}_{LO} &= \sigma_{\rm a} {c} B^{(n+1)} -\sigma_{\rm a} c E^{(n+1)}_{LO} \label{eqn:rade_eq}\\
\frac{{\bf F}^{(n+1)}_{LO}-{\bf F}^*_{LO}}{c\Delta t} + \nabla \cdot \mathbb{f}^{(n+1)}cE_{LO}^{(n+1)} &= -\sigma_{\rm t} {\bf F}^{(n+1)}_{LO}  \label{eqn:radF_eq}
\end{align}
In practice, the cell-averaged radiation flux is obtained from the surrounding face-normal fluxes as:
\begin{align}
    {F}_{LO,i,x} = \sum_j \mathbf{n}_{ijx} \cdot {\bf F}_{LO,ij} /\sum_j |{\mathbf{n}_{ijx}}| ,   \\
    {F}_{LO,i,y} = \sum_j \mathbf{n}_{ijy} \cdot {\bf F}_{LO,ij} /\sum_j |{\mathbf{n}_{ijy}}| ,   
\end{align}
Face averaged radiation energy density $E_{ij}$ is computed using  arithmetic averaging: 
\begin{align}
    E_{LO,ij} =\frac{ E_{LO,i} + E_{LO,j} }{2}
\end{align}
For the HO sector, MMC terms are included via the cell-averaged source terms $D_E$ and $D_F$:
\begin{align}
    D_{E,i,\nu} &= -\sigma_{t,\nu} \frac{\mathbf{v}_i}{{c}}\cdot \left(\mathbf{F}^{(n+1)}_{LO,i,\nu} - \frac{4}{3} E^{(n+1)}_{LO,i,\nu} \mathbf{v}_i \right),\\ \nonumber
    D_{F,i,\nu}  &= \boldsymbol{\Omega} \cdot \mathbf{v}_i E^{(n+1)}_{LO,i,\nu},
\end{align}
These terms, $D_{E,i,\nu}$ and $D_{F,i,\nu}$, contribute to the cell-averaged source term $S_i$ in Eq.\ \ref{eq:moc_soln}. The spatial discretization of the  MMC terms is spatially consistent between LO and HO systems, ensuring compatibility across the radiation hydrodynamics framework.

\subsection{Initial and boundary conditions}

The application of initial conditions in the MOC sector uses a particle based interpolation, following the procedure described in Section~\ref{sec:moc_disc} for the application of $I_0$ at the beginning of each time step.
For the first time step of the simulation the initial intensity in the domain is set to a Planck function at a given initial temperature, $T_0$.
For non-uniform initial temperature, the vertex-based initial intensities use the temperature in the cell that they would travel back in time into, so the general form of the initial intensity of a particle in 2D is:
\begin{align}
    I_0 = \frac{c}{4\pi}\left(B_{\nu}(T_0^{--})\Theta(\Omega_1>0)\Theta(\Omega_2>0) + B_{\nu}(T_0^{-+})\Theta(\Omega_1>0)\Theta(\Omega_2<0) \right.\nonumber\\\left. + B_{\nu}(T_0^{+-})\Theta(\Omega_1<0)\Theta(\Omega_2>0) + B_{\nu}(T_0^{++})\Theta(\Omega_1<0)\Theta(\Omega_2<0)\right) \;\;,
\end{align}
where $T_0^{\pm\pm}$ are the temperatures in the cells, $\Omega_k$ is the component of direction along dimension index $k$, and $\Theta(\cdot)$ is the unit step function.
The LO system uses the same $T_0$ values to initialize the LO cell-centered radiation energy density,
\begin{align}
    E_{LO,0} = B(T_0) = aT_0^4 \;\;.
\end{align}
We permit the initial material temperature to be a different value, $T_{m,0}$, permitting initialization of problems out of equilibrium.

We discretize the Miften-Larsen boundary condition as
\begin{align}
    F_{LO,j} =
    \left(F_{HO,j}^+ + F_{HO,j}^-\right)\frac{E_{LO,i}}{E_{HO,i}} - 2F_{HO,j}^- \;\;,
\end{align}
where subscript $i$ indicates cell-centered evaluation and subscript $j$ indicates evaluation at a face of cell $i$.
To use the vertex-based HO samples, we linearly average the partial fluxes evaluated at the two (four) vertices bounding face $j$ (cell $i$).
We note that the discrepancy between evaluation at $i$ versus $j$ incurs spatial truncation error in the boundary condition (though one may reconstruct $E_{LO,j}$).

\subsection{Domain decomposition and adaptive mesh refinement}
\label{sec:dlbamr}

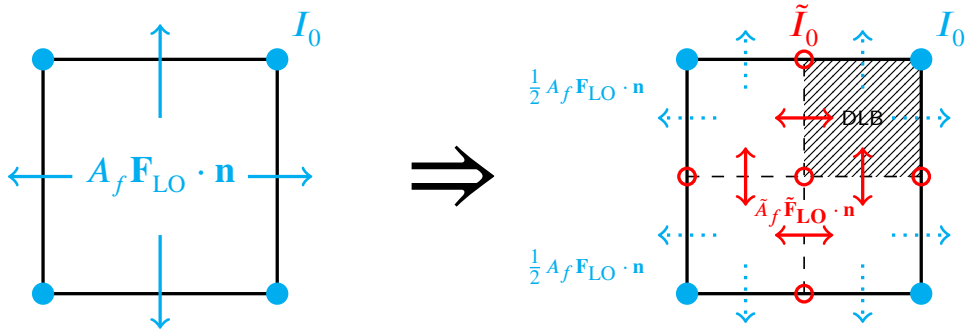
\begin{figure}[h!]
    \centering
    \begin{minipage}{0.8\textwidth}
        \centering
        \resizebox{\textwidth}{!}{
\begin{tikzpicture}


\draw[black,thick] (0,0) rectangle (2,2);
\filldraw[cyan] (0,0) circle (2.5pt); 
\filldraw[cyan] (2,0) circle (2.5pt); 
\filldraw[cyan] (0,2) circle (2.5pt); 
\filldraw[cyan] (2,2) circle (2.5pt); 

\node[above right, cyan] at (2,2) {$I_0$};

\draw[->, thick, cyan] (1,1.5) -- (1,2.3);    
\draw[->, thick, cyan] (1.75,1) -- (2.3,1);    
\draw[->, thick, cyan] (1,0.5) -- (1,-0.3);   
\draw[->, thick, cyan] (0.25,1) -- (-0.3,1);   

\node[cyan] at (1,1) {$A_f{\bf F}_{\rm LO}\cdot {\bf n}$};

\node at (3.5,1) {\Huge$\Rightarrow$};

\begin{scope}[xshift=5.5cm]

  \draw[thick] (0,0) rectangle (2,2);
  
  \fill[pattern=north east lines] (1,1) rectangle (2,2);
  \node[black] at (1.5,1.5) {\tiny DLB};
  
  \filldraw[cyan] (0,0) circle (2.5pt); 
  \filldraw[cyan] (2,0) circle (2.5pt); 
  \filldraw[cyan] (0,2) circle (2.5pt); 
  \filldraw[cyan] (2,2) circle (2.5pt); 

  \node[above right, cyan] at (2,2) {$I_0$};
  
  \draw[dashed] (1,0) -- (1,0.5);
  \draw[dashed] (1,0.8) -- (1,2);
  \draw[dashed] (0,1) -- (2,1);

  \draw[thick, red] (1,0) circle (2pt);
  \draw[thick, red] (1,1) circle (2pt);
  \draw[thick, red] (1,2) circle (2pt);
  \draw[thick, red] (0,1) circle (2pt);
  \draw[thick, red] (2,1) circle (2pt);

  \node[above, red] at (1,2) {$\tilde{I}_0$};

  \draw[<->, red, thick] (0.5,0.75) -- (0.5,1.25);
  \draw[<->, red, thick] (0.75,0.5) -- (1.25,0.5);
  \draw[<->, red, thick] (1.5,0.75) -- (1.5,1.25);
  \draw[<->, red, thick] (0.75,1.5) -- (1.25,1.5);

  \node[red] at (1,0.7) {\tiny $\tilde{A}_f\bf\tilde{F}_{LO}\cdot{\bf n}$};
  
  \draw[->, cyan, thick, dotted] (0.25,0.5) -- (-0.25,0.5);
  \draw[->, cyan, thick, dotted] (1.75,0.5) -- (2.25,0.5);
  \draw[->, cyan, thick, dotted] (0.25,1.5) -- (-0.25,1.5);
  \draw[->, cyan, thick, dotted] (1.75,1.5) -- (2.25,1.5);
  \draw[->, cyan, thick, dotted] (0.5,0.25) -- (0.5,-0.25);
  \draw[->, cyan, thick, dotted] (0.5,1.75) -- (0.5,2.25);
  \draw[->, cyan, thick, dotted] (1.5,0.25) -- (1.5,-0.25);
  \draw[->, cyan, thick, dotted] (1.5,1.75) -- (1.5,2.25);

  \node[above left, cyan] at (-0.25,1.5) {\tiny $\frac{1}{2}A_f{\bf F}_{\rm LO}\cdot {\bf n}$};
  \node[below left, cyan] at (-0.25,0.5) {\tiny $\frac{1}{2}A_f{\bf F}_{\rm LO}\cdot {\bf n}$};
\end{scope}

\end{tikzpicture}
}
\end{minipage}
\caption{Diagram of AMR and DLB for HO and LO systems in a spatial cell undergoing one level of refinement (dashed gray lines) and the upper right cell being sent to another MPI rank (hatched area).
The blue solid nodes and arrows represent the original vertex-based MOC/HO intensity and face-based LO normal flux, respectively.
The red nodes and arrows represent the new MOC intensity particles and interior normal fluxes, respectively, obtained from interpolation.
The blue dotted arrows represent the partitioned LO normal flux from the original cell, determined from old-new face area ratios (in a first pass) and continuity corrections (in a second pass).
To evaluate refinement in the hatched area, the original blue arrow and node data is needed on the rank containing the hatched area.}
\label{fig:amr}
\end{figure}

Our implementation is compatible with dynamic load balancing (DLB) of mesh cells between MPI ranks and cell-based adaptive mesh refinement (AMR).
Furthermore, we permit multiple levels of mesh refinement and coarsening between time steps.
These optimizations are important to the implementation of our method with multidimensional hydrodynamics.
Details discussed here are diagrammed in 2D in Figure \ref{fig:amr} for one cell undergoing one level of refinement to four cells, with on refined cell being sent to another MPI rank.

We track both DLB and AMR changes with maps of old to new and new to old cell indices. 
For the LO system, we store flux times face area per face per cell (flux at faces dotted with the outward cell normal), and restrict these through cell coarsening by simply summing the values along the outer edges of the cell cluster that has been removed.
For prolongation through cell refinement, at the old coarse edges, orthogonal to the flux (and linear interpolation) direction, we prolongate to new refined faces (whose union is the old coarse face) by multiplying the old coarse flux by the fraction of area the new refined face has, so the sum of the flux over the refined faces is the old flux of the coarse face.
To obtain fluxes on new faces interior to the old cell, we linearly interpolate per-area fluxes along each dimension between the edges of the old coarse cell (multiplying by the refined face areas for the new persistent values).
Continuity is automatically obtained on the interior of the refined cell, but needs to be separately enforced at the boundary of a newly refined cell adjacent to a previously refined region.

For the MOC particles, we first leverage the exiting IMC particle remap capability in Jayenne to map existing MOC particles through DLB.
However, unlike IMC, particles do not represent conserved entities so can be spawned and removed through AMR.
Particle removal (or restriction) simply uses the cell maps to identify which old refined cell clusters have changed into new coarse cells, and removes the particles on the interior of the new coarse cell.
In principle this removal step is not needed, since the census will be replaced and the particles that would be removed by this procedure should be orphaned by possessing invalid vertex indices in the new mesh. 
We spawn (or prolongate) new particles at new vertices in refined cells using bilinear spatial interpolation in 2D, separately for each direction ordinate; each particle stores a multigroup intensity, so the interpolation uses four multigroup vectors from the four particles bounding the old coarse cell to obtain one particle at a new interior vertex.
We note that this procedure may spawn duplicate MOC particles at new vertices on cell edges, if two adjacent cells have each been refined, but the intensity of the duplicates should be consistent due to the continuity of intensity at vertices.
Nonetheless, it may be useful to remove duplicates to minimize memory usage in the particle census.

Spawning particles in a domain-decomposed DLB setting may require old particles be duplicated across MPI ranks, since multiple MPI ranks may need the old particle in prolongation procedures on refined cells adjacent to the old vertex.
Consequently, after an initial pass of the IMC-like particle remap, we use the old-new cell maps to build additional communication buffers for particles needed for prolongation/spawning.
This involves each MPI rank building a particle request list for particles it needs, participating in an all-all gather of the request per rank, creating a buffer per rank to send particles it has to ranks that need the particle, inverting this communication pattern so receiving ranks know the size of buffers to expect from each other rank, then performing the MPI send and receives.
A similar procedure for fluxes at the faces of coarse old cells is needed, as multiple ranks may need the old value at a particular face.

\subsection{Algorithm outline}
\label{sec:algo}

The full MOC algorithm is outlined in Algorithm~\ref{fig:algo}.
Within a time step, the MOC particle transport takes place in an outer loop, and an inner implicit iterative LO solver uses the closure data from the outer loop.
In turn, this LO data resets the source terms for particle transport in the next outer iteration. 
In this outline, the $||\cdot||_k$ $k$-norms are evaluated over spatial cells, and furthermore vector notation $\vec{\cdot}$ implies a vector over spatial cell index.
The vector notation is introduced in order to merely indicate that the equation for each cell, used to evaluate the Jacobian, is a function of radiation energy density over other cells.
The information content of the outline focuses on the specifics of the MOC radiative transfer solution, and omits some formulae discussed in the prior sections, for brevity.

\begin{algorithm}[!h]
    \caption{Outline of MOC-LO solution algorithm}\label{fig:algo}
    \begin{algorithmic}
    \State Initialize intensity $I_0$, energy density $E_{LO}^{(n=0)}$, flux $\mathbf{F}_{LO}^{(n=0)}$ and material temperature $T_0$.
    \For{time step $n = 0$ to $n_{\max}-1$}
    \State Do AMR/DLB and hydrodynamic solve.
    \State Evaluate cell-group opacity $\sigma_{a,g}$, $\sigma_{s,g}$, cell-face-group opacity $\sigma_{t,g}$.
    \State Prolong+restrict/remap $I_n$, $\mathbf{F}_{LO}^{(n)}$ (Section \ref{sec:dlbamr}).
    \State Calculate $E_{LO}^*$ and flux $\mathbf{F}_{LO}^*$ using explicit LO MMC corrections (Eqs. \eqref{eqn:lommc1}).
    \State Initialize outer iteration: $E_{LO}^{(k=0)} = E_{LO}^*$, $\mathbf{F}_{LO}^{(k=0)} = \mathbf{F}_{LO}^*$, $T^{(k=0)} = T_n$.
    \State Set {\tt all\_done} = false.
    \While{not {\tt all\_done}}
    \State Source MOC particles at $t_p = t_n+\Delta t_n = t_{n+1}$, at mesh vertices.
    \For{each MOC particle $p$}
    \State Transport position across cell in direction $-\mathbf{\Omega}_p$ (backward in time).
    \State Integrate each $\tau_{p,g}$, $I_{p,g}$ over cell step (Eq. \eqref{eq:moc_recurse}).
    \If{$t_p = t_n$ or $\tau_p>\tau_{\rm threshold}$}
    \State Stop transport of particle $p$.
    \State Apply final intensity update to particle if $t_p=t_n$ (Eq. \eqref{eq:moc_apply_ic}).
    \State Send $p$ to original vertex coordinate at $t_{n+1}$.
    \EndIf
    \EndFor
    \State Calculate closure coefficients $A^{(k+1)}$, $\mathbb{B}^{(k+1)}$, $\mathbb{X}^{(k+1)}$ from particles, $I^{(k+1)}$ (Eqs. \eqref{eq:gcfdt_diag}, \eqref{eq:gcfdt_cros}).
    \State Calculate $E_{HO,g}^{(k+1)}$-weighted absorption opacity, $\sigma_a$.
    \State Initialize inner iteration: $E_{LO}^{(l=0)}=E_{LO}^{(k)}$, $\mathbf{F}_{LO}^{(l=0)} = \mathbf{F}_{LO}^{(k)}$.
    \State Set {\tt done} = false
    \While{not done}
    \State Use Newton-Raphson method to solve $\rho c_v(T^{(l)}-T_n)/\Delta t_n=c\sigma_aE_{LO}^{(l)}-\sigma_Pca(T^{(l)})^4$ for $T^{(l)}$.
    \State Set $r_l = \mathtt{res}(E_{LO}^{(l)}, \mathbf{F}_{LO}^{(l)}, T^{(l)})$
    \State Evaluate LO cell-cell Jacobian matrix $\mathcal{J}_{i,i'}^{(l)} = \left.\frac{\partial g_i}{\partial E_{LO,i'}}\right|_l$, for LO system $g_i(\vec{E}_{LO}^{(l)},\vec{T}^{(l)}) = 0$.
    \State Solve $\vec{g} = \mathcal{J}^{(l)}(\vec{E}_{LO}^{(l+1)} - \vec{E}_{LO}^{(l)})$ for $\vec{E}_{LO}^{(l+1)}$.
    \State Evaluate $\mathbf{F}_{LO}^{(l+1)}$ with $\mathbf{F}_{LO}^*$, $E_{LO}^{(l+1)}$, $A^{(k+1)}$, $\mathbb{B}^{(k+1)}$, $\mathbb{X}^{(k+1)}$.
    \If{l > 0}
        \State {\tt done} = $r_{l} / r_{l-1} < \varepsilon'$
    \EndIf
    \State $l \gets l+1$
    \EndWhile
    \State Set $E_{LO}^{(k+1)}=E_{LO}^{(l+1)}$, $\mathbf{F}_{LO}^{(k+1)} = \mathbf{F}_{LO}^{(l+1)}$, $T^{(k+1)} = T^{(l+1)}$.
    \If{k > 0}
        \State {\tt all\_done} $||(E_{LO}^{(k+1)}-E_{LO}^{(k)})/E_{LO}^{(k+1)}||_{\infty} > \varepsilon$
    \EndIf
    \State $k \gets k+1$
    \EndWhile
    \State Set $E_{LO}^{(n+1)} = E_{LO}^{(k)}$, $\mathbf{F}_{LO}^{(n+1)} = \mathbf{F}_{LO}^{(k)}$, $T_{n+1} = T^{(k)}$
    \State Calculate energy-momentum coupling terms.
    \EndFor
    \Procedure{$\mathtt{res}$}{$E_{LO}$, $\mathbf{F}_{LO}$, $T$} \\
    \Return $||(E_{LO} - E_{LO,n})/\Delta t_n + \nabla\cdot\mathbf{F}_{LO} + \sigma_a cE_{LO}-\sigma_PcaT^4||_2$
    \EndProcedure
    \end{algorithmic}
\end{algorithm}

\section{Test problems}\label{sec:test}

\subsection{Plane-parallel vacuum}

Here we test the MOC and low-order system together on a 1D vacuum (0 opacity) problem \citep{RyanDolence2020}.
The problem consists of a plane-parallel region with a isotropic boundary condition.
The condition of 0 opacity makes the problem a stress test of ray effects in the high-order MOC
sector (or any method employing a discrete angular quadrature) and causal signal speed in the low-order sector.
The problem has an analytic solution for the intensity,
\begin{align}
    I(x,t) = I_0\Theta(ct - x/\mu) \;\;,
\end{align}
where $x$ is the spatial coordinate, $t$ is time, $\mu$ is the component of direction along $x$,
$I_0$ is the isotropic intensity value at the boundary, and $\Theta(\cdot)$ is the unit step function.
Thus the radiation energy density a linear function in space,
\begin{align}
    E = \frac{1}{c}I_0\left(1 - \frac{x}{ct}\right) \;\;,\;\; x\leq ct \;\;.
\end{align}
Incorporating the analytic solution for intensity into the Eddington tensor gives the following for
the diagonal tensor element for the dimension along the 1D coordinate,
\begin{align}
    f_{xx} = \frac{1}{3}\left(1 + \frac{x}{ct} + \left(\frac{x}{ct}\right)^2\right)
    \;\;,\;\; x\leq ct \;\;.
\end{align}
By noting that the diagonal of the Eddington tensor must sum to 1, we obtain the diagonal entries
for the ($y$-)directions transverse to the $x$-direction,
\begin{align}
    f_{yy} = \frac{1}{2}(1 - f_{xx}) \;\;.
\end{align}
The off-diagonal values, $f_{xy}$, are analytically 0, which the selected angular quadrature
obtains to machine precision.

A generalization of the linear formula for radiation energy density can be obtained for
a purely absorbing medium (ignoring re-emission),
\begin{align}
    E(x,t) = \frac{1}{c}I_0\left(E_2(\sigma_ax) - \frac{x}{ct}E_2(\sigma_act)\right)
    \;\;,\;\; x\leq ct \;\;,
    \label{eqn:erad_sol_nonvac}
\end{align}
where $E_2(\cdot)$ is the exponential integral function of the 2nd kind.

\begin{figure}[h!]
    \centering
    \includegraphics[width=0.48\textwidth]{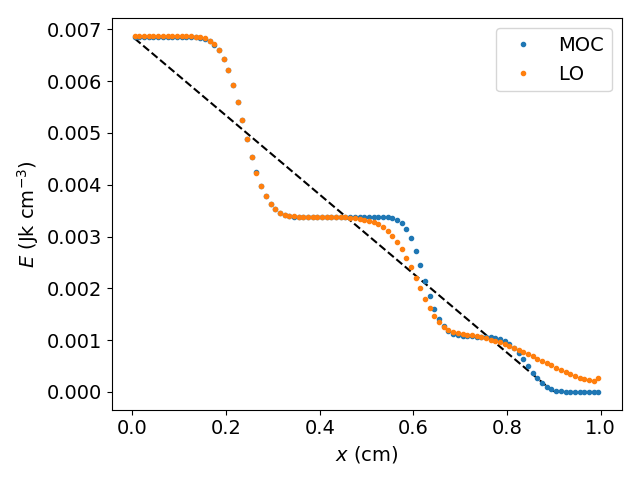}
    \includegraphics[width=0.48\textwidth]{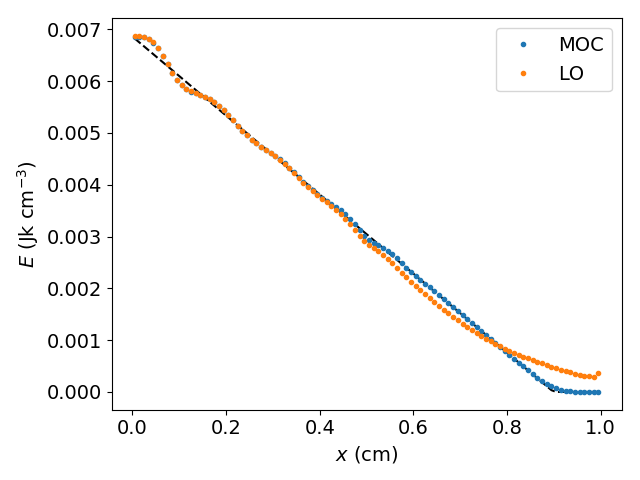} 
    \caption{Radiation energy density versus position for the plane-parallel vacuum problem at $t=0.003$ sh.
    The analytic solution (black dashed) is compared to the MOC (blue) and LO (orange) solutions.
    Left panel: triangular Chebyshev-Legendre discrete ordinate quadrature of order 6.
    Right panel: the same quadrature but of order 24.
    The LO solution is not strictly causal.
    We also observe that the LO solution is strongly influenced by ray effects from the MOC sector.}
    \label{fig:vac_erad_s24} 
\end{figure}

We consider a surface source corresponding to 1 keV, a 1 cm domain, and a final time of 0.003 sh.
In Fig.~\ref{fig:vac_erad_s24}, we use 100 uniform spatial cells and 50 uniform time steps.
Consequently, the time step size is about 1.8 times the light crossing time of a cell.
For directional quadrature, we use a triangular Chebyshev-Legendre quadrature set, available in
the Draco component library for radiative transfer\footnote{https://github.com/lanl/Draco}.
This quadrature has the polar symmetry axis aligned in a direction orthogonal to our 2D mesh, hence
the results shown here are symmetric under reflection of the $x$ and $y$ coordinates.

\begin{figure}[h!]
    \centering
    \includegraphics[width=0.48\textwidth]{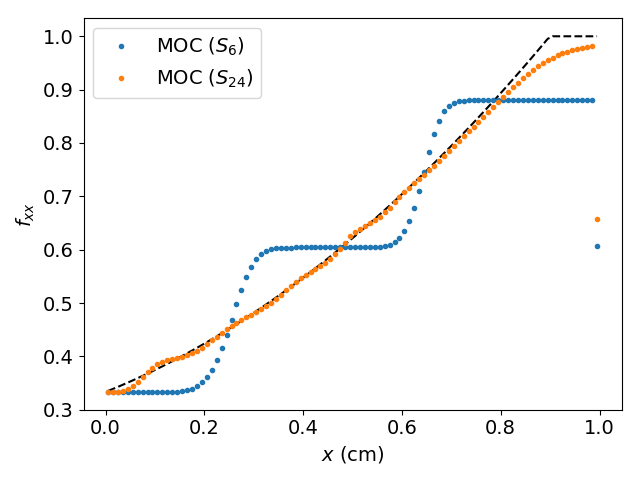}
    \includegraphics[width=0.48\textwidth]{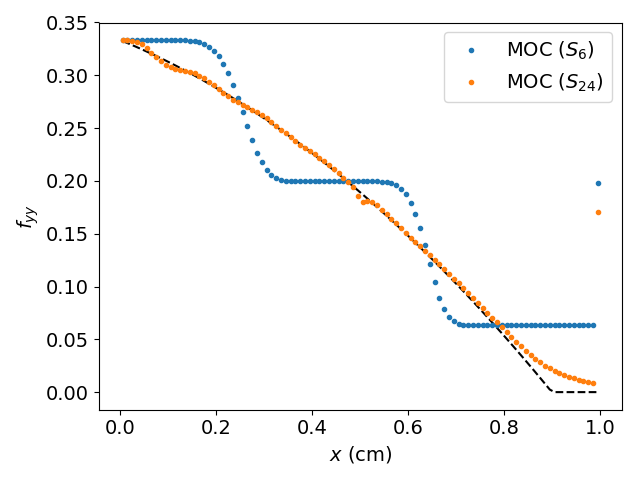} 
    \caption{Eddington tensor components versus position for the plane-parallel vacuum problem at $t=0.003$ sh.
    The analytic solution (black dashed) is compared to MOC solutions with triangular Chebyshev-Legendre 
    quadratures of order 6 (blue) and order 24 (orange).
    Left panel: the diagonal $xx$-component.
    Right panel: the diagonal $yy$-component.}
    \label{fig:vac_edd_tens} 
\end{figure}

The diagonal components of the Eddington tensor versus the $x$-coordinate are shown in
Fig.~\ref{fig:vac_edd_tens}, again for triangular Chebyshev-Legendre quadratures of order 6 (blue)
and order 24 (orange).
Similar to the radiation energy density, we can see ray effects in space in the Eddington tensor at
low quadrature order.
We see the Eddington tensor does not reach the analytic value at the causal limit of the wavefront.
This is due in part to the presence of an initial small ambient radiation field in the problem (added
to avoid singular Eddington tensors past the causal wavefront).

\begin{figure}[h!]
    \centering
    \includegraphics[width=0.48\textwidth]{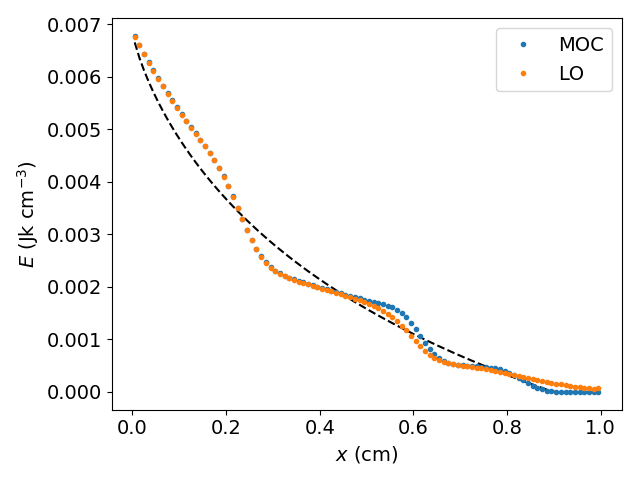}
    \includegraphics[width=0.48\textwidth]{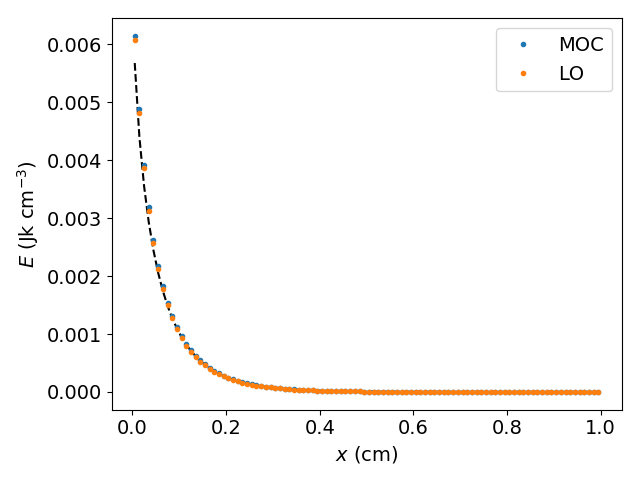} 
    \caption{Radiation energy density versus position for the plane-parallel non-vacuum pure-absorbing problem
    at $t=0.003$ sh.
    The analytic solution (black dashed), given by~\eqref{eqn:erad_sol_nonvac}, is compared to MOC (blue)
    and LO (orange) solutions with triangular Chebyshev-Legendre quadrature of order 6.
    Left panel: absorption opacity $\sigma_a=1$ cm$^{-1}$, which shows prominent ray effects.
    Right panel: absorption opacity $\sigma_a=10$ cm$^{-1}$, which has ray effects suppressed.}
    \label{fig:nonvac_erad} 
\end{figure}

Figure~\ref{fig:nonvac_erad} has radiation energy density versus the $x$-coordinate for non-zero absorption
opacity, $\sigma_a=1$ cm$^{-1}$ (left panel) and $\sigma_a=10$ cm$^{-1}$ (right panel), at $t=0.003$ sh.
The spatial and temporal discretization is the same as the preceding vacuum problem.
The order 6 quadrature shows mitigated ray effects persist at $\sigma_a=1$ cm$^{-1}$, but are not apparent
at $\sigma_a=10$ cm$^{-1}$.


\subsection{Optically thick scattering pulse}

The evolution of a Gaussian pulse of radiation energy in an optically thick purely scattering medium centered on the origin has an analytic time-dependent solution \citep{McKinney+2014}:
\begin{align}
    E\left({\bf x}, t\right) = E\left({\bf x}, t_0\right) \exp \left( \frac{-3 \sigma_{\rm s} \lVert {\bf x} \rVert^2}{4 c \left( t + t_0 \right)} \right) \left( \frac{t+t_0}{t_0}\right)^{-n/2}
\end{align}
where $n$ is the dimensionality of the problem and $t_0$ is the initial time. 

This problem tests several key features of numerical transport schemes: elastic scattering terms on both the low order and high order sector, and asymptotic preserving fluxes in the optically thick regime. 
For our method, this test can also be used to measure the agreement between moment and MOC representations of the radiation energy density, which should be driven to consistency by the scattering source term in the MOC transport equation. 
Here we consider diffusion in 1D ($n = 1$) and set $\sigma_{\rm s} = 1000~{\rm cm}^{-1}$, $t_0 = 0.025{\rm ~sh}$, $t_{\rm f} = 0.0125{\rm ~sh}$, and $T_{\rm r,0} = 1 {\rm ~keV}$, on a 2D domain of size $1~{\rm cm}$ in each direction (with half as many cells in the symmetric $y$ dimension as in $x$ in all cases).
We use a base timestep of $0.00025~{\rm sh}$.

\begin{figure}[h!]
    \centering
    \includegraphics[width=\textwidth]{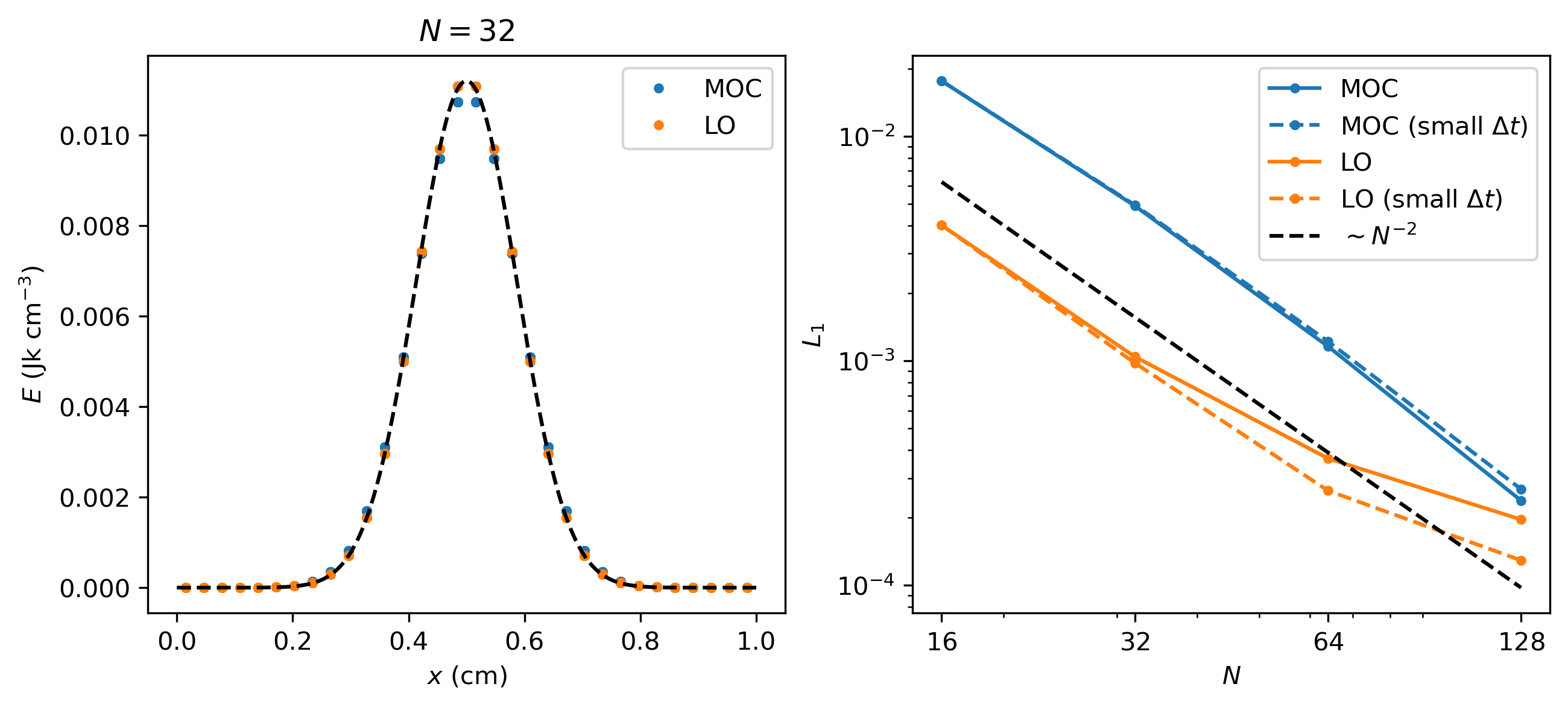} 
    \caption{Diffusing Gaussian test problem. 
    In all cases the MOC result is compared against the analytic solution at $t_{\rm f} = 0.0125{\rm ~sh}$. 
    The left panel shows the result (both ``true" low-order energy, and non-conservative MOC energy density derived from a direct integral of the samples) for $N=32$ cells along the $x$ axis. 
    The right panel shows the rate of convergence relative to $N^{-2}$ for both MOC and LO representations. 
    The solid lines have a fixed timestep of $\Delta t = 0.00025~{\rm sh}$, whereas the small $\Delta t$ runs have a timestep reduced proportionally as we increase resolution. 
    In the former case, the LO solution begins to converge at first order at high resolution due to our first order accuracy in time.}
    \label{fig:gaussian_static} 
\end{figure}

Figure \ref{fig:gaussian_static} shows the results of our method relative to the analytic solution for a range of resolutions. 
MOC and LO representations of the radiation energy are in good agreement, and our scheme shows second order spatial convergence when the timestep is appropriately reduced in proportion to the increasing spatial resolution.
Evidently our method effectively captures the diffusion limit in a pure scattering medium.

\begin{figure}[h!]
    \centering
    \includegraphics[width=\textwidth]{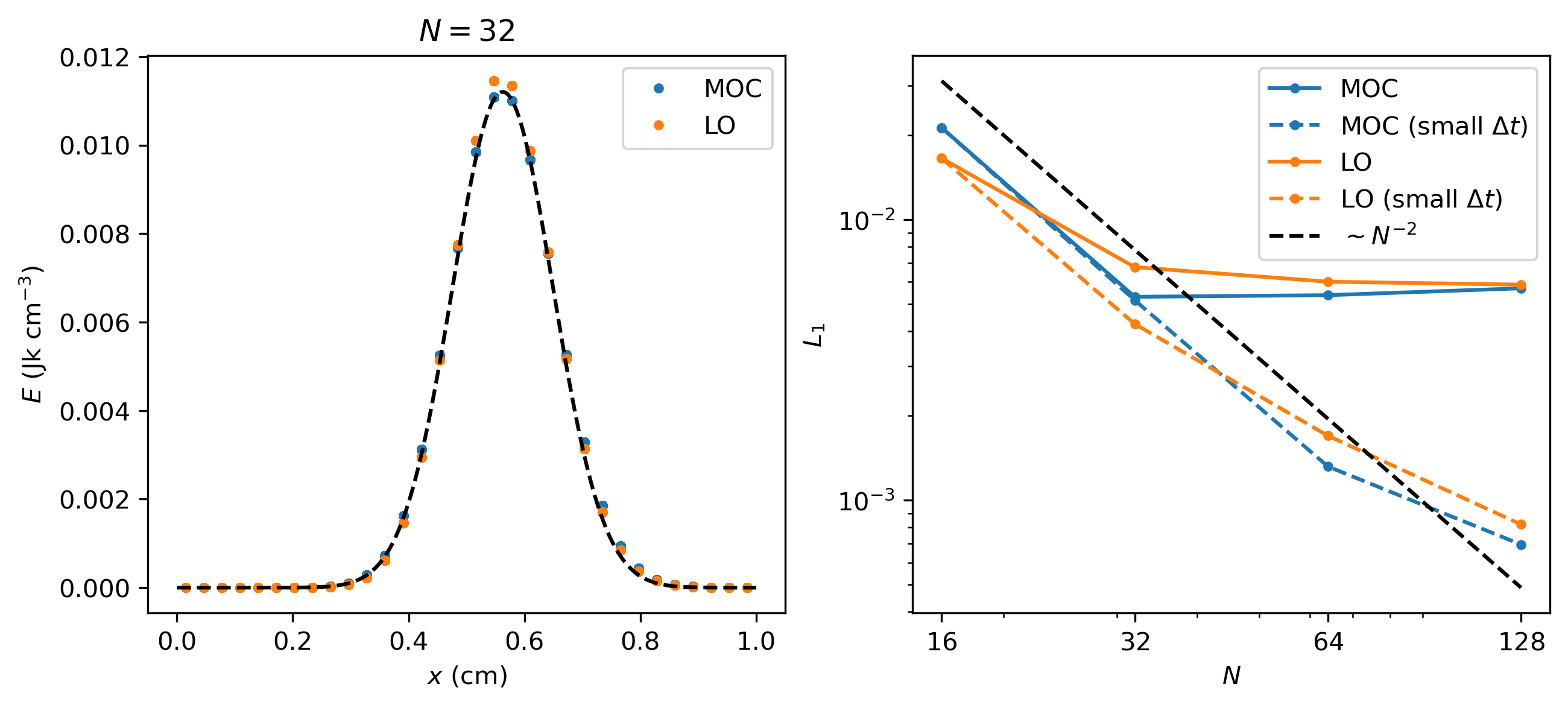} 
    \caption{Diffusing Gaussian test problem boosted with velocity $0.017c {\bf \hat{x}}$. 
    In all cases the MOC and LO (true) results are compared against the analytic solution at $t_{\rm f} = 0.0125{\rm ~sh}$. 
    The left panel shows the result for $N=32$ cells along the $x$ axis, where the pulse has advected along the $x$ axis.  
    The right panel shows the rate of convergence relative to $N^{-2}$ for both MOC and LO representations. 
    Convergence degrades more severely than in the static case when the timestep is held constant.}
    \label{fig:gaussian_boost} 
\end{figure}

We can additionally apply a boost to this problem such that the pulse advects along the grid while diffusing. 
This further tests our implementation of moving material corrections. Here, we choose a boost velocity of $0.017c$ in the $+x$ direction. 
Figure \ref{fig:gaussian_boost} shows the result of this test for a fixed timestep of $0.00025~{\rm sh}$. 
Convergence is lost entirely when the timestep is held fixed; when the timestep is decreased in proportion to the cell size, convergence is substantially recovered, although apparently at less than second order at the highest spatial resolutions. 
The \cite{LowrieWollaber2014} formalism does not formally converge with $\mathcal{O}\left( \beta \right)$, and this may contribute to the observed error.

\subsection{Marshak wave}
\label{sec:mwave}

A standard family of tests of thermal-radiative coupling and the diffusion limit, depending on optical
thickness, are Marshak waves, which typically involve a fixed surface source at one boundary inducing a
wave into an initially uniform medium.
For all forms of the Marshak wave problem examined here, we use a MOC optical depth threshold for inverse
time tracking of $\tau=16$.
We test the MOC scheme on a Marshak wave with an isotropic surface source of $T_s=$1 keV, uniform initial
volume (radiation and material) temperature of 1 eV, a heat capacity of 0.3 GJ/keV/cm$^3$, and an absorption
opacity satisfying
\begin{align}
    \sigma_{a,\nu} = \sigma_a = 30\left(\frac{T_s}{T}\right)^3 \;\;\text{cm}^{-1} \;\;.
\end{align}
We use a domain of 1 cm in length along the $x$-coordinate, and a time range of 1 sh.
The ratio of maximum to minimum possible optical thickness per spatial cell is $10^9$, which for the
length of the domain will span very optically thick to thin regimes.

\begin{figure}[h!]
    \centering
    \includegraphics[width=0.48\textwidth]{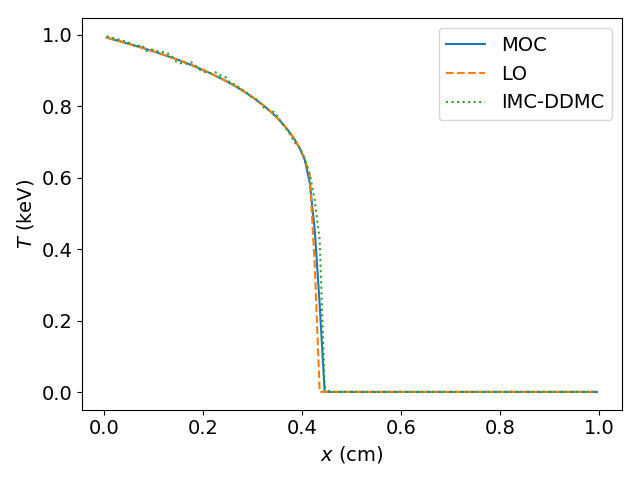} 
    \caption{Radiation temperature versus $x$-coordinate for the Marshak wave test.}
    \label{fig:marshak} 
\end{figure}

\begin{figure}[h!]
    \centering
    \includegraphics[width=0.48\textwidth]{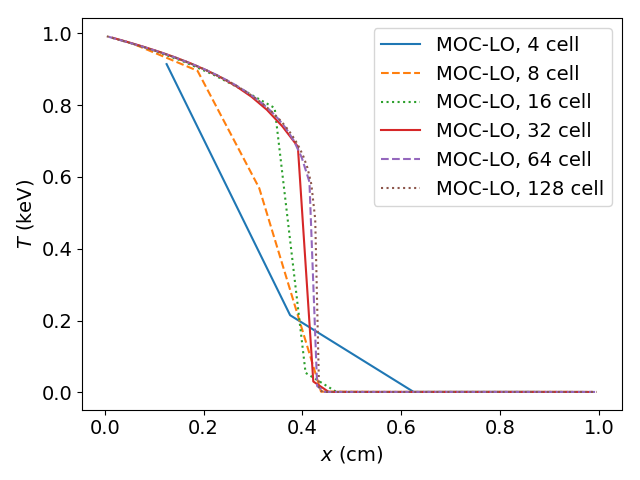}
    \includegraphics[width=0.48\textwidth]{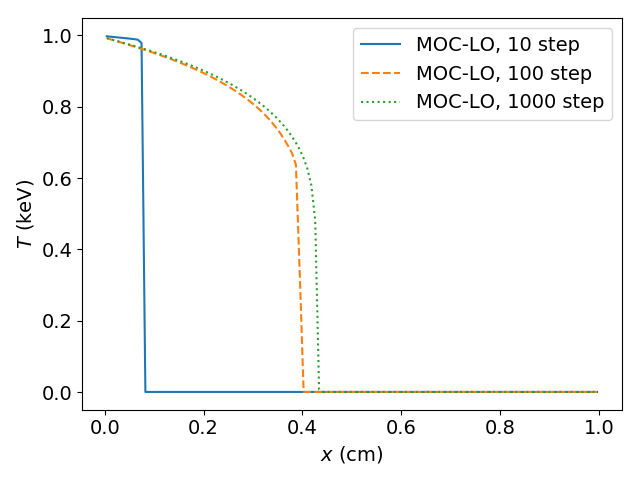}    
    \caption{The left panel shows material temperature versus $x$-coordinate for the Marshak wave test,
    for different spatial resolutions.
    The right panel is the same as the left, but for different time resolutions.}
    \label{fig:marshak_res} 
\end{figure}

Testing this Marshak wave problem at different spatial resolutions numerically demonstrates adherence to
the asymptotic diffusion limit, as can be seen in the left panel of Fig.~\ref{fig:marshak_res}, where each
curve corresponds to a number of uniform spatial cells across the 1 cm domain.
Similarly, testing this problem at different temporal resolutions shows adherence to the maximum principle
\citep{Larsen+1987b}, where each curve corresponds to a number of uniform time steps over the 1 sh span.

\begin{figure}[h!]
    \centering
    \includegraphics[width=0.48\textwidth]{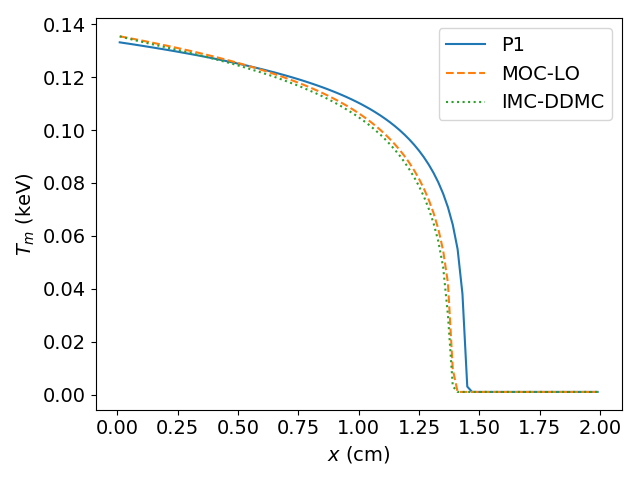}
    \includegraphics[width=0.48\textwidth]{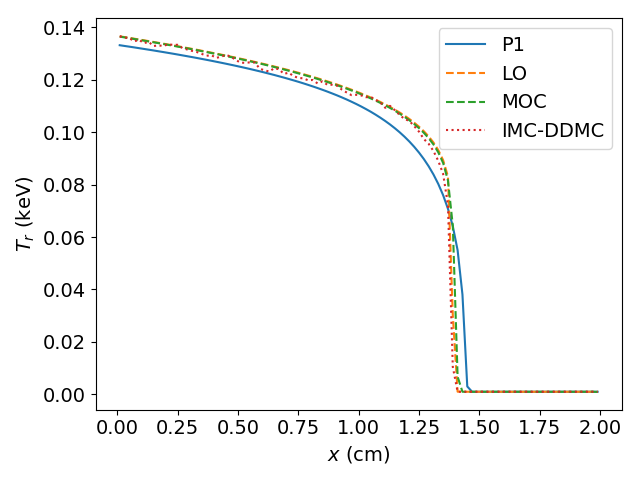}    
    \caption{The left panel shows material temperature versus $x$-coordinate at 5 sh
    for the thin Marshak wave test for P1 (blue solid), MOC (orange dashed), and IMC-DDMC (green dotted).
    The right panel is the same as the left, but for radiation temperature, and with
    LO (orange dashed) and MOC (green dashed) sectors separated (IMC here is red dashed curve).}
    \label{fig:marshak_thin} 
\end{figure}

We now test the optically thin regime with the same problem, but a surface source of $T_s=0.15$ keV, a
heat capacity of 0.013784 GJ/keV/cm$^3$, and an absorption opacity
\begin{align}
    \sigma_{a,\nu} = \sigma_a = 10^{-3}\left(\frac{1}{T}\right)^3  \;\;\text{cm}^{-1} \;\;,
\end{align}
where $T$ is in keV.
Furthermore, we use 100 uniform spatial cells over 2 cm, 5000 uniform time steps over 5 sh, and for MOC
we employ a $S_6$ triangular Chebyshev-Legendre quadrature.
Results for $P_1$, MOC, and IMC-DDMC are shown in Fig.~\ref{fig:marshak_thin} for material (left panel)
and radiation temperature versus the spatial coordinate at 5 sh.
In this thin Marshak wave test, we see that radiation temperature for MOC and IMC-DDMC having a steeper
wavefront than the corresponding material temperature solutions.
We also see that $P_1$ does not achieve the correct wave speed.

\begin{figure}[h!]
    \centering
    \includegraphics[width=0.48\textwidth]{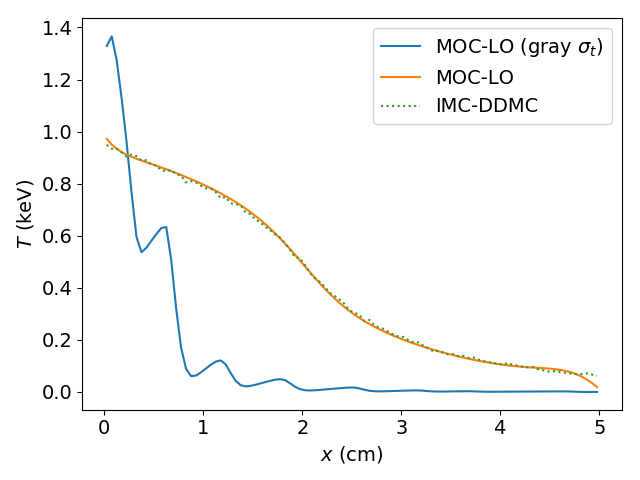}
    \includegraphics[width=0.48\textwidth]{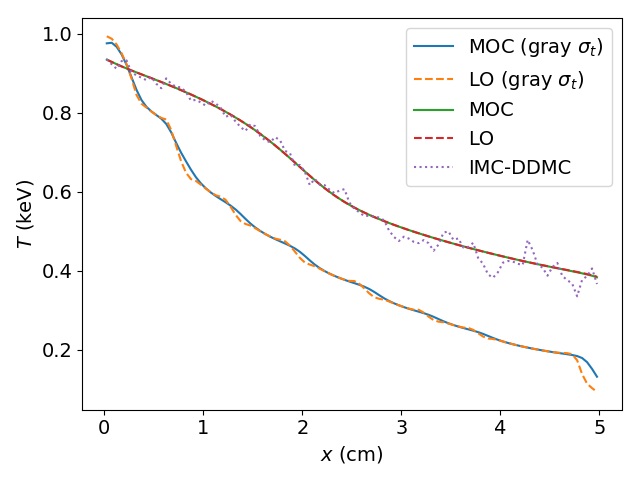}    
    \caption{The left panel shows material temperature versus $x$-coordinate at 5 sh
    for the multifrequency Marshak wave test for MOC-LO using gray average total face opacities ($\sigma_t$)
    (blue solid), MOC-LO using the $\Delta t$-dependent coefficients (solid) and IMC-DDMC (green dotted).
    The right panel is the same as the left, but for radiation temperature, and with
    gray-$\sigma_t$ LO (orange dashed) and MOC (blue solid) sectors separated, as well as
    the LO (red dashed) and MOC (green solid) sectors spearated (IMC here is purple dashed curve).}
    \label{fig:mg_marshak}
\end{figure}

We finally test a multifrequency form of the Marshak wave problem, with
\begin{align}
    \sigma_{a,\nu} = 30\left(\frac{1}{T}\right)^{1/2}\left(\frac{1}{\nu}\right)^3  \;\;\text{cm}^{-1} \;\;,
\end{align}
where $T$ and $\nu$ are in keV.
Scalar multiples of this opacity profile are used by~\cite{Densmore+2012} to test multifrequency
behavior.
The heat capacity is set to 0.3 GJ/keV/cm$^3$.
We run this problem with 100 uniform spatial cells over 5 cm, 1000 uniform time steps over 1 sh duration,
and 32 logarithmic frequency groups from 0.01 eV to 100 keV.
We again use an optical depth threshold for inverse particle tracking of $\tau=16$, and again use an
$S_6$ triangular Chebyshev-Legendre quadrature; it is worth emphasizing that the optical depth threshold is compared
conservatively to the group with the lowest optical depth aggregate.
To demonstrate the importance of properly accounting for multifrequency behavior, in addition to IMC-DDMC
we compare to a form of the MOC-LO method where the total face opacity is first averaged over group, then
the LO system is time-discretized.
This alternate approach may represent a naive formulation, without a middle order or consistency term,
which would support better matching the HO multigroup flux.
Figure~\ref{fig:mg_marshak} has material (left panel) and radiation (right panel) temperature versus
space for the multifrequency Marshak wave, for MOC, IMC-DDMC and MOC with the alternate face opacity
treatment, labeled MOC (gray $\sigma_t$) or MOC-LO (gray $\sigma_t$).
Evidently, the alternate naive approach fails to properly capture the HO flux behavior, resulting in a non-physical profile and maximum principle violation in the material temperature.
In contrast, the MOC-LO solution using the coefficients from Section~\ref{sec:moments} satisfies the maximum principle, and matches very closely with the IMC-DDMC solution in both radiation and material temperatures.
We note however that the naive approach uses an arithmetic flux-weighted multigroup average of $\sigma_t$; harmonically averaging may provide a better result.

\subsection{Mach 45 shock}

Material motion corrections, including momentum coupling, are important in matching the semi-analytic
Mach 45 radiative shock solution given by~\cite{LowrieWollaber2014}, based on derivation of
\cite{LowrieEdwards2008} (see also references therein).
This is a standard benchmark for radiation-hydrodynamics codes: incorrect accounting for
momentum coupling or the bulk effects of Doppler shift can cause significant deviation from
the semi-analytic solution.

\begin{figure}[h!]
    \centering
    \includegraphics[width=0.48\linewidth]{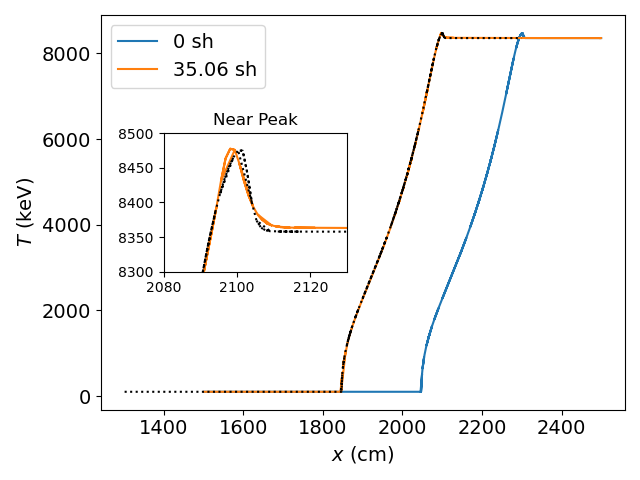}
    \includegraphics[width=0.48\linewidth]{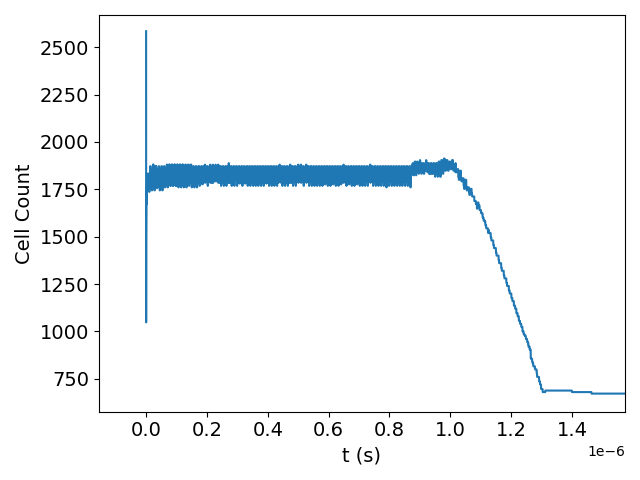}
    \caption{The left panel shows material temperature versus $x$-coordinate at 35.06 sh
    for the Mach 45 radiative shock test, for MOC-LO (orange solid).
    Also plotted are the initial profile (blue) and the initial profile displaced by
    the distance traveled by the shock speed after 35.06 sh (black dotted).
    The inset is zoomed in near the peak temperature.
    The right panel shows total spatial cell count versus time for the same simulation.}
    \label{fig:mach45}
\end{figure}

For this problem, we use a triangular Chebyshev-Legendre quadrature at order $S_4$ and
an optical depth threshold $\tau=16$.
We use this problem to exercise 3-level AMR with in planar geometry, with the coarsest level
corresponding to 200 uniform cells in the $x$-direction from 1500 to 2500 cm, and 2 cells in
the transverse $y$-direction (thus 400 cells total if the domain is at the coarsest resolution
everywhere).
Figure~\ref{fig:mach45} has material temperature versus $x$-coordinate at 35.06 sh (left panel)
and cell count versus time (right panel).
Along with the MOC-LO solution, the semi-analytic solution that serves as the initial condition is
displaced in space by the shock speed times 35.06 sh (black dotted line).
We see good agreement of MOC-LO with the semi-analytic solution in the overall structure of the
shock front at 35.06 sh, although the inset shows a slight discrepancy in shape of the peak.
The cell count is seen to immediately jump and stabilize around 1800 cells, indicating that
there is a rough conservation of cell resolution as the shock traverses the mesh.
The cell count starts to drop at 100 sh, when the shock is displaced $\sim$500 cm and hence
starting to leave the simulated spatial domain.
Since the spatial mesh is 2D, this problem also tests the ability of the AMR to preserve the
1D profile in the transerse direction, as refinement increases the cells in that direction as
well.

\subsection{Crooked pipe}
\label{sec:cpipe}

We next run the ``crooked pipe'' or ``top hat'' test (see \cite{Gentile2001}, \cite{jiang_godunov_2012}).  We test in Cartesian coordinates, making our solutions most directly comparable to \cite{jiang_godunov_2012}.  Compared to the setup in that paper, our domain is halved along the axis of symmetry and the coordinate axes are inverted — thus we solve over the domain $(0, 2) \times (0, 7)$ cm.  To form square cells, we use 100 $\times$ 350 cells in total, each 0.02 cm to a side — slightly higher resolution in $x$ but lower in $y$ than the 128 $\times$ 512 cells over the comparable domain in \cite{jiang_godunov_2012}.  Reflecting boundaries were used at the $x = 0$ boundary. The pipe geometry, the material densities and opacities, and the source temperature remain exactly similar to \cite{jiang_godunov_2012}.  The timestep was set to $10^{-11}$s, or the light-crossing time of five zones.

\begin{figure}[h!]
    \centering
    \includegraphics[width=0.7\textwidth]{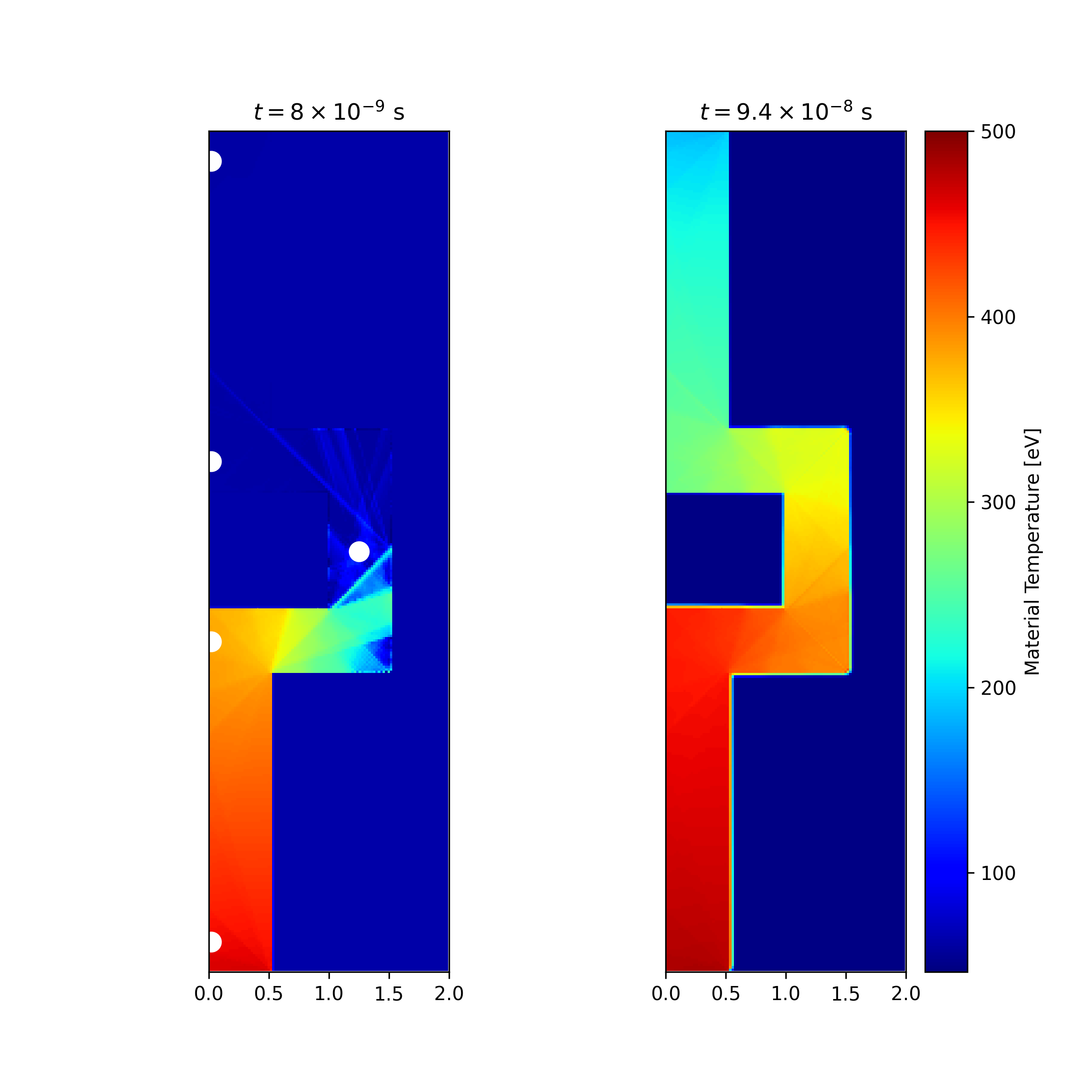}
    \caption{Temperature in the crooked pipe as solved by MOC, at $8\times10^{-9}$s and $9.4\times10^{-8}$s.  White points indicate the temperature probes measured in the next figure.} 
    \label{fig:crooked_pipes} 
\end{figure}

\begin{figure}[h!]
    \centering
    \includegraphics[width=0.95\textwidth]{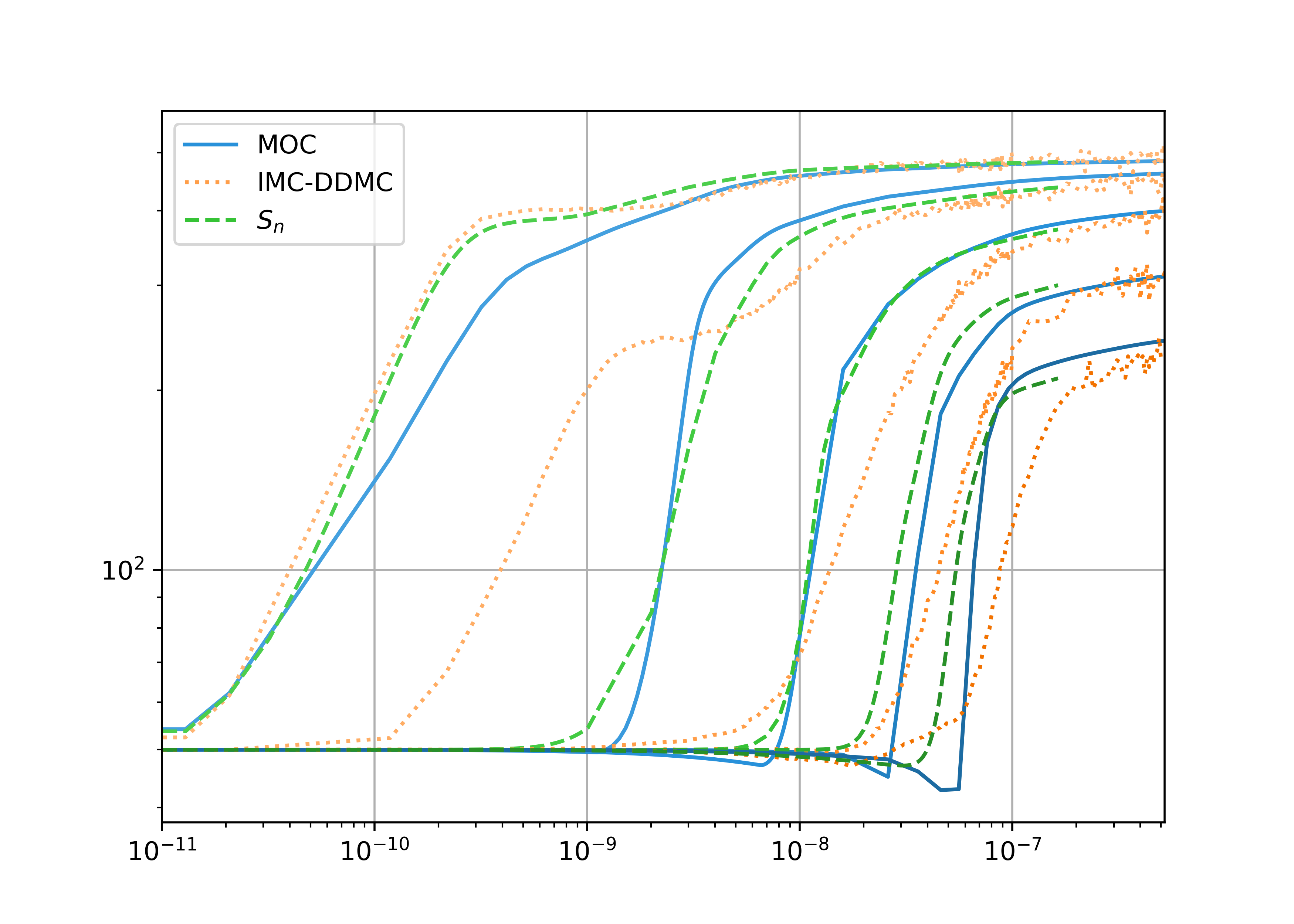}
    \caption{Radiation temperature versus time at probe points indicated in Figure \ref{fig:crooked_pipes}, run with three methods: MOC, IMC-DDMC, and S$_n$.  All points are overplotted, with darker lines indicating later probes.  MOC qualitatively matches IMC and S$_n$ arrival times and heating rates.  Remaining differences are resolution-dependent: very low spatial and temporal resolution cause the MOC solution to move more slowly, higher resolutions cause slightly earlier and faster heating.}
    \label{fig:crooked_methods} 
\end{figure}

Figure \ref{fig:crooked_pipes} shows the material temperature in a simulation using MOC at time $8\times10^{-9}$s, and later at $9.4\times10^{-8}$s.  Radiation from a uniform hot source at $y = 0$ first heats the central baffle, which re-emits and heats the outer wall, and so on until radiation reaches and heats the later straight section of pipe.
The material temperature is measured at each of the indicated dots, presented in Figure \ref{fig:crooked_methods}. Placement of these probes is exactly similar to \cite{jiang_godunov_2012} — in our coordinates, $(0, 0.25),\,(0, 2.75),\,(1.25, 3.5),\,(0, 4.25),$ and $(0, 6.75)$, measuring two points directly heated by the source, one within the baffle, and two after.  Note this placement puts probes 1, 2, 4, and 5 against the reflecting boundary at the left edge of the domain, but the boundary condition does not have an appreciable effect on the result — probes at $x = 0.1$ produced visually indistinguishable results.

Figure \ref{fig:crooked_methods} compares heating at each probe when the identical problem is run using IMC-DDMC, S$_n$, and MOC transport. The simulation with IMC-DDMC used 400,000 particles, constant throughout the simulation. S$_n$ was run with sixth-order quadrature, 
and MOC was run with sixth-order level-symmetric quadrature.
These parameters were chosen to represent the best-case converged behavior, looking for unavoidable method biases rather than evaluating convergence or relative performance for constant accuracy. In particular, the test is quite sensitive to differences in the weighting scheme when interpolating face-centered opacities to cell centers -- only harmonically averaged opacities produced the correct initial heating times of later probes.

\subsection{Coax}
\label{sec:coax}

\begin{figure}[h!]
    \centering
    \includegraphics[width=0.8\textwidth]{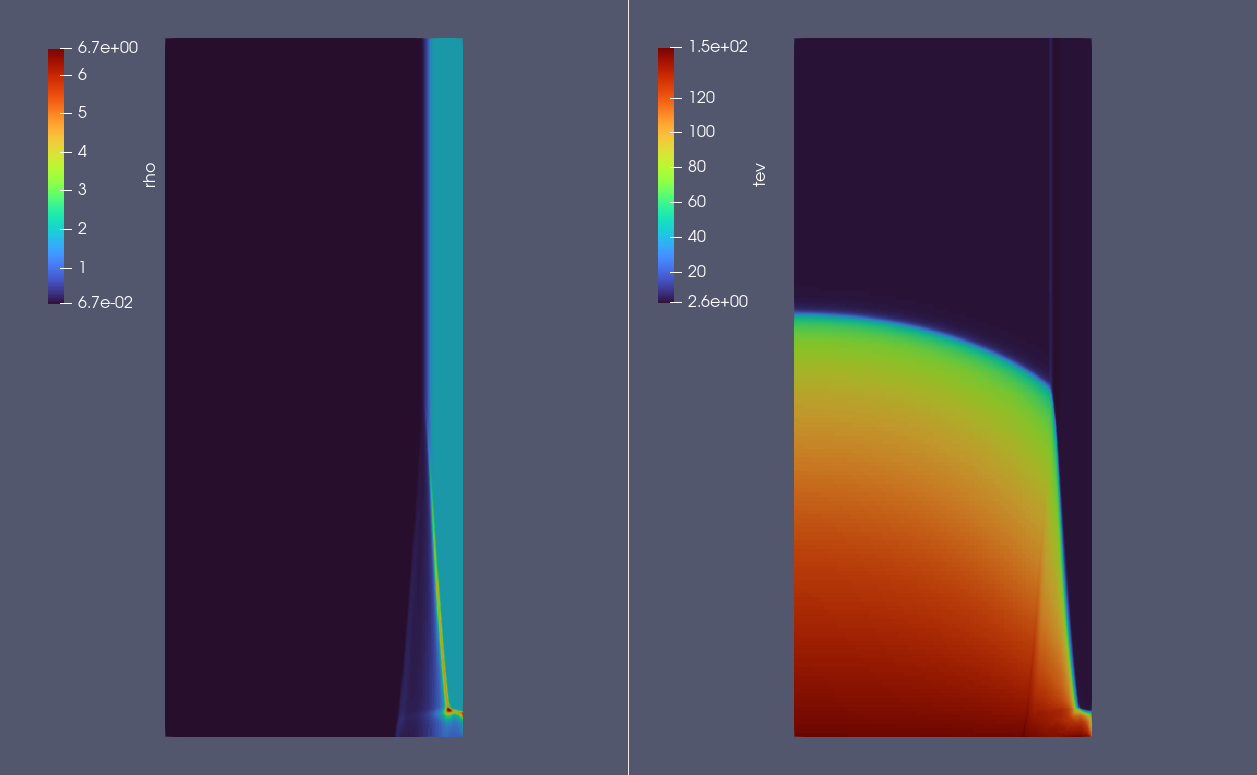} 
    \caption{Contour of density (left) and material temperature (right). }
    \label{fig:coax_pic} 
\end{figure}

\begin{figure}[h!]
    \centering
    \includegraphics[width=0.8\textwidth]{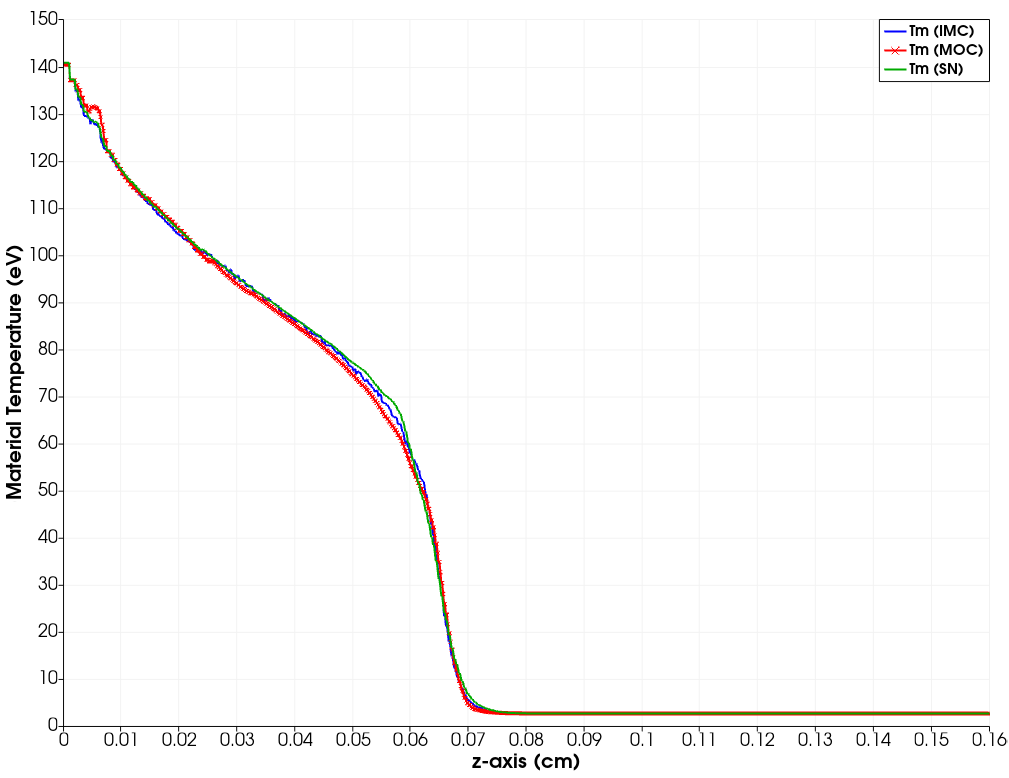} 
    \caption{Comparison of material temperatures at r=0.06cm, t=1.0ns. }
    \label{fig:coax_lineout} 
\end{figure}

Next, we apply our methods to a multifrequency radiation-hydrodynamics test problem in cylindrical geometry, in which radiation flows through a pipe ($\rho=1.845[g/cc]$) lined with a foam ($\rho=0.070[g/cc]$) \citep{Fryer+2020}. Beryllium opacity is used for the pipe material. The system is initially in equilibrium at T=4eV, and driven by an isotropic 150 eV radiation source at z=0.00 cm. The pipe's inner diameter and thickness are 0.12 cm and 0.016 cm, respectively, and the frequency domain was divided into 55 groups. Fig.\ \ref{fig:coax_pic} presents contour plots of the density and material temperature at $t=1.0ns$, while Fig.\ \ref{fig:coax_lineout} compares the material temperature along the original interface ($r=0.06cm$), as computed by MOC and Sn (both using $S_6$ level symmetric quadrature) and by IMC (with approximately 5 million particles per time step). All three approaches exhibit excellent agreement.


For these three method implementations, we also compare Coax runtime performance using 50 MPI ranks each, on Intel Xeon Platinum 8479 CPUs.
The number of particles for IMC-DDMC is set to 10 million total, and particles are apportioned to cells according to their surface source and emission energy.
Load balancing for IMC-DDMC is done either by number of spatial cells or by particle work per rank (particle-based dynamic load balancing, PDLB.
PDLB is decoupled from the hydrodynamic decomposition with ParMETIS \cite{Karypis2003}, described in detail by for the Jayenne IMC-DDMC implementation by \cite{Long2023}.
MOC and S$_n$ are again set to use S$_6$ level-symmetric quadrature.
The results are given in Table \ref{tab:coax}, with the total runtime over the number of time steps is shown to 0.1 and 1.0 ns of simulation time.
\begin{table}[h!]
    \centering
    \begin{tabularx}{\textwidth}{X|X|X|X|X}
         Stop time (ns) & IMC-DDMC+PDLB & IMC-DDMC & S$_n$ & MOC \\
         \hline\hline
         0.1 & 88 / 72 & 403 / 72 & 252 / 125 & 240 / 72 \\
         1.0 & 835 / 252 & 1499 / 252 & 1079 / 353 & 2326 / 252
    \end{tabularx}
    \caption{Runtimes (in seconds) for each method over total number of time steps, using 50 MPI ranks.
    Each row corresponds to a physical stop time for the simulation, in nanoseconds.}
    \label{tab:coax}
\end{table}

We see that MOC, as implemented, is the least performant of the method implementations over the 1.0 ns of time, approximately twice as slow.
As expected, MOC and IMC-DDMC+PDLB runtimes scale approximately with the amount of physical time simulated, given particle trajectory lengths scale with time, and S$_n$ runtime scales with the number of time steps.
Furthermore, we can also see that PDLB aids IMC-DDMC significantly at early time, where IMC-DDMC is poorly load balanced over ranks \citep{Long2023}.
S$_n$ takes more time steps due to a library-internal time step controller not active for the other method implementations; moreover over one third of the time steps are within the first 0.1 ns.

We note that applying the optical depth threshold does not speed up MOC for this problem (Section \ref{sec:tau}), due to the conservative application of the threshold in determining particle termination with multigroup opacity.
However, these initial comparisons are promising, considering that aside from software optimization (for instance, GPU porting) there is a possibility of further performance gains from spatially or angularly adaptive quadratures.

\subsection{Optical depth threshold tests}
\label{sec:tau}

Finally, we consider the effect of the optical depth threshold on MOC run time performance.
The speed-up relative to $\tau=\infty$ for a 100 time step crooked pipe (as given in Section \ref{sec:cpipe}), a 32 time step Coax (as given in Section \ref{sec:coax}), and a 1000 time step Marshak wave (first version given in Section \ref{sec:mwave}) simulation are shown in Table \ref{tab:tau_test} for three different optical depth threshold values.
For these comparisons all tests use an $S_6$ level-symmetric quadrature.
All simulations were run on CPUs, using domain decomposition for multi-core runs: 112 cores for crooked pipe, 64 cores for Coax, and 1 core for Marshak.
These comparisons are dependent on software implementation details, and may not represent a fully optimized implementation or relative performance on GPUs. 
\begin{table}[h!]
    \centering
    \begin{tabularx}{0.7\textwidth}{X|X|X|X}
         $\tau$ & Crooked pipe & Coax & Marshak wave \\
         \hline\hline
         16 & 1.27 & 1.00 & 1.66 \\
         32 & 1.27 & 0.99 & 1.66 \\
         64 & 1.23 & 1.00 & 1.66 \\
         $10^9$ & - & - & 1.53
    \end{tabularx}
    \caption{MOC runtime speed-up for optical depth thresholds $\tau=16,32,64$ relative to $\tau=\infty$, to two significant decimal places.
    All simulations use $S_6$ level-symmetric angular quadratures.}
    \label{tab:tau_test}
\end{table}

For the crooked pipe simulations, we note that there is not significant difference in speed between the optical depth thresholds tested; this is consistent with the optical depth increasing rapidly relative to the number of cells traversed in the thick wall, as MOC particles travel backwards.
The optical depth of the wall cells is 40 mean-free paths, so transport beyond these cells is consistent with the slight drop in performance from $\tau=32$ to $\tau=64$.

The run time performance of Coax problem is evidently insensitive to the optical depth threshold: no significant change in performance was observed (each of the 4 simulations were run 3 times to mitigate run-to-run noise in the comparisons).
Since the Coax problem is multigroup, and MOC particles are only terminated when the lowest optical depth associated with the particle exceeds the threshold, this insensitivity is consistent with optically thin groups causing particles to reach the beginning of each time step.

The grey Marshak wave simulations exhibit similar behavior to the crooked pipe simulations, for the optical depth thresholds tested, where we only observe significant performance gain in comparison to the infinite optical depth simulation.
We additionally test the problem with an optical depth threshold of $10^9$, which for 128 spatial cells corresponds to 4 cell traversals per particle at the minimum temperature, hence the only optical depth threshold with non-trivial transport work in cells at the minimum temperature.
We see that around or above this optical depth threshold, the speed-up is affected.

\section{Conclusion}\label{sec:conc}


We have developed a method for thermal radiation transport using a method of long characteristics (MOC) that traces particles backward in time, within a time step.
The backward motion of the particles importantly permits us to terminate the particle rays at an optical depth threshold for optimization, and to precisely set the intensity solution points.
We have set the solution points to be the vertices of the spatial mesh, permitting angle and frequency integrals without reconstruction at vertices, and simple averaging on faces, and in cells.
Furthermore, we have coupled the high-order MOC solution with an implicit gray low-order (LO) solution that enforces energy conservation and couples to matter.
The LO solution has been interfaced to multigroup MOC using coefficients summed over frequency groups: time has been discretized before the sum, making these coefficients resemble flux- or Eddington tensor-weighted harmonic averages of the number of mean-free times per time step.
The MOC intensity solution in turn uses the LO radiation energy density and flux in source terms representing scattering and material motion corrections.

Additionally, in part by leveraging existing software infrastructure, we have made our MOC and LO solutions compatible with radiation-hydrodynamics, adaptive mesh refinement (AMR), and dynamic load balancing (DLB).
AMR and DLB entail the implementation of standard prolongation/restriction operations and MPI send/receives, respectively, for the LO flux solution.
For the MOC particle census representing the intensity, we have supplemented standard particle communication (as is typically used for Monte Carlo particles) with particle-spawning from interpolated intensity, forming a type of particle-based prolongation operation.

We have applied the method to a standard battery of numerical tests for thermal radiation transport and radiation hydrodynamics.
Our results for each test are summarized as follows.
\begin{itemize}
    \item The \textbf{plane-parallel vacuum} test demonstrates that our method suffers ray effects for standard angular quadratures, as expected. We also observe the effect of non-strict causality due to the implicitness of the LO system of equations, inducing non-zero intensity beyond the wavefront.
    \item The \textbf{optically thick scattering pulse} demonstrates that our method reproduces the standard spreading Gaussian shape of the radiation energy density. We advect the pulse to test the material motion corrections as well. This test shows second order spatial convergence when the time step is reduced in proportion to the spatial cell size. Convergence can be lost entirely (saturated) when the time step is held fixed and the cell size is reduced.
    \item A gray optically thick, gray optically thin, and multifrequency form of the \textbf{Marshak wave} test shows that our method agrees closely with IMC-DDMC at sufficiently high resolution. The optically thick test shows that our method and implementation does not suffer from teleportation error or maximum principle violations. The optically thin test shows that our method achieves better accuracy than a simple $P_1$ closure. The multifrequency test shows that our multigroup formulation using time step-dependent closures achieves good accuracy relative to an IMC-DDMC solution; we have also demonstrated very poor accuracy with a naive gray LO solution.
    \item The \textbf{Mach 45} non-equilibrium gray radiative shock test demonstrates that our method can reproduce the semi-analytic solution, which consists of the steady-state shock form advecting at the shock speed. Given that the solution is sensitive the momentum coupling and material motion corrections, this test shows our treatment of these terms is correct. With AMR, we can also see that our MOC and LO prolongation-restriction operations near the wavefront do not significantly degrade the quality of the solution.
    \item The \textbf{crooked pipe} (or ``top hat'') test shows that our method functions in 2D planar geometry. At the same space and time resolutions, MOC performs comparably to IMC-DDMC and S$_n$ in terms of wave speed through the pipe, indicated by temperature  at a standard set of tracer points. However, we note that the MOC-LO solution is sensitive to how the face opacities are evaluated at the thick-thin interface; we found that we need harmonic face opacity averaging for accurate wall heating.
    \item The \textbf{Coax} problem is the most sophisticated test in terms of physics fidelity; it includes realistic multifrequency (multigroup) opacity, cylindrical geometry, multidimensional hydrodynamics, AMR and DLB. It is also a test that we have found is sensitive to excluding the time step-dependent closures, both for diagonal and off-diagonal Eddington tensor contributions to the flux. Stated differently, we find that these closures furnish agreement to IMC-DDMC and S$_n$, relative to naive gray closures. The resulting agreement is excellent between the three methods (or as good as we could anticipate).
\end{itemize}
With the Marshak, crooked pipe and Coax tests we have explored the impact of optimization via the optical depth threshold, and find $\sim$20\% and $\sim$66\% speed-up in the thick Marshak and crooked pipe tests relative to using no optical depth threshold.
For Coax, we observed no speed-up, consistent with each particle termination being determined by the optically thinnest group, hence not activating prior to the particle reaching the beginning of the time step.
We also note that our MOC implementation is two to three times slower than IMC-DDMC and S$_n$ on the Coax problem, for reasonable input parameters for these methods.

Our plan for future work is to enable spatially adaptive angular quadratures, permitting further optimization in regions where the intensity is isotropic or low, or otherwise unimportant to the goals of a simulation.
We also anticipate exploring angularly adaptive quadratures, where intensity and other information may inform where to apply higher resolution on the direction unit sphere.
Angular adaptivity can be useful in particular for problems like the plane-parallel vaccum, where the intensity becomes highly anisotropic as one moves away from the surface source.
Additionally, we plan to examine the use of bilinear finite elements for integrating along MOC rays and bilinear reconstruction for LO temperature, to reduce spatial truncation error.
In line with this, we intend to revisit the face opacity averaging, as there may be more sophisticated or less heuristic averaging approaches that are readily applicable (for instance using asymptotic boundary layer analysis, \cite{Habetler1975,Densmore+2007}, or more information from MOC).

Other items to consider for future work involve physics fidelity, for instance: relativistically accurate material motion corrections (see, for instance, \cite{Lindquist1966,Shibata2011,RyanDolence2020,Lowrie2023,Laiu2025}) and more sophisticated scattering models (for instance, for Compton scattering, as explored by \cite{Till2020,Mcgraw2023}).
Finally, it may be worth considering how to treat the MOC-LO discretization with tensor trains for compression and further optimization, which has recently been explored for S$_n$ by \cite{Gorodetsky+2025}.
Given the long characteristics and LO system, tensorizing our approach would presumably entail different requirements than S$_n$.

\section{Acknowledgments}\label{sec:ack}

It is a pleasure to thank M. Cleveland and J. Warsa for useful discussions. This work has been assigned document release number LA-UR-25-29734. This work was supported by the U.S. Department of Energy through the Los Alamos National Laboratory. Los Alamos National Laboratory is operated by Triad National Security, LLC, for the National Nuclear Security Administration of U.S. Department of Energy (Contract No. 89233218CNA000001).


\printcredits

\bibliographystyle{unsrt}

\bibliography{manuscript-refs}



\end{document}